\documentclass[twocolumn,aps,prd,tighten,amssymb,amsmath,nofootinbib]{revtex4-1}
\usepackage{amssymb}
\usepackage{epsfig}
\usepackage{amsmath}
\usepackage[]{algpseudocode}

\usepackage{color}
\usepackage{xcolor}
\usepackage{graphicx}

\newcommand{\trK}{\mathrm{tr}K}
\newcommand{\trT}{\mathrm{tr}T}
\newcommand{\trKhat}{\mathrm{tr}{\hat K}}

\begin{document}

\title{A Simflowny-based finite-difference code for high-performance computing in Relativity}

\author{
Carlos Palenzuela$^{1,2}$,
Borja Mi\~{n}ano$^{2}$,
Daniele Vigan\`o$^{1,2}$,
Antoni Arbona$^{2}$,
Carles Bona-Casas$^{1,2}$,
Andreu Rigo$^{2}$,
Miguel Bezares$^{1,2}$,
Carles Bona$^{1,2}$,
Joan Mass\'o$^{1,2}$
}

\affiliation{${^1}$Departament  de  F\'{\i}sica,
 Universitat  de  les  Illes  Balears  and  Institut  d'Estudis
Espacials  de  Catalunya,  Palma  de  Mallorca,  Baleares  E-07122,  Spain}
\affiliation{$^{2}$ Institut Aplicacions Computationals (IAC3)  Universitat  de  les  Illes  Balears,  Palma  de  Mallorca,  Baleares  E-07122,  Spain}%

\begin{abstract}
The tremendous challenge of comparing our theoretical models with the gravitational-wave observations in the new era of multimessenger astronomy requires accurate and fast numerical simulations of complicated physical systems described by the Einstein and the matter equations. These requirements can only be satisfied if the simulations can be parallelized efficiently on a large number of processors and advanced computational strategies are adopted.
To achieve this goal we have developed {\it Simflowny}, an open platform for scientific dynamical models which automatically generates parallel code for different simulation frameworks,
 allowing the use of HPC infrastructures to non-specialist scientists. One of these frameworks is SAMRAI, a mature patch-based structured adaptive mesh refinement infrastructure, 
capable of reaching exascale in some specific problems. Here we present the numerical techniques that we have implemented on this framework by using {\it Simflowny} in order to perform fast, efficient, accurate and highly-scalable simulations. These techniques involve high-order schemes for smooth and non-smooth solutions, Adaptive Mesh Refinement with arbitrary resolution ratios and an optimal strategy for the sub-cycling in time. We  validate the automatically generated codes for the SAMRAI infrastructure with some simple test examples (i.e., wave equation and Newtonian MHD) and finally with the Einstein equations. 
\end{abstract}

\maketitle

\section{Introduction}

The recent direct detections of gravitational waves (GWs), consistent with the emission by binary black-hole mergers~\citep{LIGO:2016blz,LIGO:2016nmj,LIGO:2017vtc,LIGO:2017a,LIGO:2017b}, have been without doubt one of the greatest scientific achievements of the decade. The most recent observation, GW170817~\citep{LIGO:2017c}, has been associated with a binary neutron star merger, thanks to a concurrent short Gamma-Ray Burst~\citep{LIGO:2017e} and a plethora of afterglow electromagnetic signals from the same source~\citep{LIGO:2017d}. This event represents the beginning of the multimessenger astronomy era, where gravitational and electromagnetic waves emitted by the same source can be correlated to extract additional information about the astrophysical systems and their underlying physics.

The potential extraction of further information from these observations relies on our ability to realistically model the astrophysical sources producing the gravitational waves. 
In this context, one of the most interesting and challenging scenarios is the merger of two compact objects. The gravitational and the electromagnetic outcome produced during a binary coalescence can only be accurately calculated by performing extremely demanding numerical simulations.
Therefore, there is a need for flexible (to include additional micro-physical processes), efficient (so that the expensive simulations can be run on nowadays' hardware) and scalable (such that the jobs can be parallelized to many processors, allowing for faster and/or more accurate simulations) codes to study numerically the emission of EM and GWs and contrast them with the observations. 

This race to High-Performance Computing (HPC)
is also present in other areas of Physics, where the common current goal is to reach exascale computing (i.e., numerical codes that can scale up to $10^6$ processors). Exascale computing would allow us, combined with modern and near-future supercomputers, to provide simulations of unprecedented size and resolution. Notice however that it is not enough to have a highly-scalable code, but it also must be fast and accurate in order to run efficiently in current clusters. The combination of these three features (speed, accuracy and scalability), the ability to switch physical models (flexibility), and the capacity to run in different infrastructures (portability) is the goal of the simulation platform {\it Simflowny}~\cite{simflowny_webpage,Arbona:2013,Arbona:2017}. {\it Simflowny} has been developed by the IAC3 group since 2008 to automatically generate complicated code and facilitate the use of HPC infrastructures to non-specialist scientists. Notice that while writing a code that scales up to a few hundred processors is within reach of many research groups, exascale is in a different league altogether. Only a few centers have the computer science experts and the capacity of developing exascale platforms. Therefore the best strategy for most science-focused groups is to leverage existing exascale projects by building specific code onto them. As we mentioned before, such code must be fast and accurate as well, and this may require profound knowledge on advanced numerical strategies to preserve the efficiency and the scalability of the selected exascale platform. This is why automatic code generation for these platforms is a sensible option. Furthermore, automatic code generation allows scientists to explore different numerical techniques and physical models in a fast and easy way, which is especially appreciated in nascent fields where models are still subject to intense toying and scrutiny.

With these requirements in mind, {\it Simflowny} was developed as an open platform for scientific dynamical models, composed by a Domain Specific Language (DSL), based on {\tt MathML} and {\tt SimML}, and a web-based integrated development environment, which automatically generates efficient parallel code for simulation frameworks. {\it Simflowny} has a simple yet ambitious goal of a complete splitting of: (i) the physical problem (i.e., the system of equations together with the initial data and the boundary conditions), (ii) the numerical methods necessary for a simulation (i.e., the discretization schemes), and (iii) the automatic generation of the simulating code, where the parallelization features of the chosen infrastructure will be optimally leveraged.
This splitting allows different types of developer profiles (physicists, computer science experts) to focus each one in their own area of expertise. In this context, Simflowny aims at maximizing such separation of concerns, including the additional separation of the physical laws (models) from the many specific problems that may use them (which include initial and boundary conditions). Other computational infrastructures also exist which allow for different degrees of flexibility, scalability and portability such as OpenModelica~\cite{openmodelica}, OpenFOAM~\cite{openfoam}, Fluidity~\cite{fluidity} and Fenics~\cite{fenics} which can be combined with Firedrake~\cite{firedrake} to provide a more complete separation of concerns. Many of them are devoted to the fluid mechanics domain. To the best of the authors' knowledge, the only platform available for the numerical relativity community is Chemora~\cite{chemora}, a partial differential equations (PDE) solving framework based on Cactus~\cite{cactus}. In this regard, Simflowny aims at eventually being able to handle not one type of evolution system (such as PDEs), but any evolution system.  Currently {\it Simflowny} can generate code for the SAMRAI infrastructure~\cite{Hornung:2002,Gunney:2016,samrai_webpage}, a patch-based Structured Adaptive Mesh Refinement Application Infrastructure developed over more than 15 years by the Center for Applied Scientific Computing at the Lawrence Livermore National Laboratory. The latest upgrades on the Adaptive Mesh Refinement (AMR) algorithms allow to improve the performance and reach a good scaling on up to 1.5 Million cores and 2 Million MPI tasks~\cite{Gunney:2016}, at least for some specific problems. 

The automatic generation of the parallelized code is obtained by writing the systems of equations and the numerical schemes in the {\it Simflowny} graphical user interface in DSL, which is then translated into the numerical language suitable for a specific infrastructure like SAMRAI. Here we  describe in detail the advanced numerical techniques that we have implemented in {\it Simflowny} in order to deal with hyperbolic-parabolic systems of equations,
which are provide as templates including:
\begin{itemize}
	\item Discretization schemes, based on the Method of Lines (MoL) which allow us to prescribe separately the discrete representation of space and time derivative operators. High-order schemes are favored, since they are more accurate and efficient at the affordable cost of limiting partially the scalability.
	\item AMR algorithm, allowing for arbitrary ratios between consecutive resolutions which enhances the scalability of the code by reducing intermediate refinement levels. This implies some modifications on the AMR schemes, with particular emphasis on the space interpolation and the sub-cycling in time strategies.
\end{itemize}
Each generated code is the {\it numerical discretization} of a problem embedded in the infrastructure. Obviously, the performance of the resulting code will depend on the specific problem, the numerical implementation and the choice of the underlying infrastructure.
Therefore, the numerical techniques above-mentioned are extremely important as it will impact on the efficiency, accuracy and scalability of the final code. 

The implementation of these numerical algorithms 
are validated by generating codes corresponding to different physical models, allowing us to analyze
their computational performance under several conditions. First we consider the scalar wave equation, that allows to compare our numerical solutions with simple analytical ones. The evaluation of the convergence of the discrete smooth solutions will be use for validating the numerical schemes and test different AMR algorithms. Secondly, we consider different Newtonian Magneto-HydroDynamic (MHD) problems, which will allow us to test the spatial discretization for non-smooth solutions as well as the AMR strategies in the presence of discontinuities and shocks. Finally, we implement the Einstein equations following the CCZ4 formulation. Several scenarios involving black holes are considered to validate the code and check its performance in weak and strong scaling tests.

\section{Discretization schemes}

The non-linear nature of different systems of equations usually requires different discretization techniques to ensure convergence and stability of the numerical solution. For instance, the wave equation can be efficiently solved by using centered finite difference discrete operators, which turn out to be unfit for solving strongly non-linear systems like the hydrodynamics equations. {\it Simflowny}, with flexibility in mind, allows for different discretization schemes, even within the same model. Here we will briefly describe the Method of Lines, which allows the combination of different spatial discretizations while keeping the same time integrator. Some commonly-used space and time discretization schemes, all of them available in {\it Simflowny}, will be discussed later.

\subsection{The Method of Lines}

Evolution equations systems of first-order in time can be written generically as
\begin{eqnarray}\label{PDEequation}
    \partial_t {\bf u} = {\cal L}({\bf u}) 
\end{eqnarray}
where ${\bf u}$ is the set of evolution fields and ${\cal L}({\bf u})$ is an operator containing arbitrary spatial derivatives of the fields. 
This system of PDEs at the continuum can be transformed into a semi-discrete problem by: (i) discretizing the spatial coordinates, $x_i = i \Delta x$ in one dimension, such that the solution is only defined in a grid of discrete points, ${\bf U}= {\bf u}(x_i)$, and (ii) substituting ${\cal L}({\bf u})$ with a discrete operator $L({\bf U})$, where the spatial partial derivatives are replaced by suitable discrete spatial derivative operators. Therefore, at each point of the grid, the continuum PDE is converted into a semi-discrete ordinary differential equation (ODE)
\begin{eqnarray}\label{semidiscrete}
    \partial_t {\bf U} = L({\bf U}) + Q_d (\bf U)
\end{eqnarray}
where $Q_d$ is an artificial dissipation operator included for stability reasons to remove high frequency modes of the solution which can not be accurately resolved in the grid. The problem can be fully discrete by defining discrete timesteps $t^n = n \Delta t$, such that the fully discrete solution at the current time can be represented as ${\bf U}^n \equiv {\bf u}(x_i, t^n)$. Explicit schemes are those for which the future solution can be written explicitly in terms of the current one, namely
\begin{equation}\label{fullydiscrete}
     {\bf U}^{n+1} = T \left[ L({ \bf U}^n  + Q_d (\bf {\bf U}^n)) \right]
\end{equation}
where $T$ can be a complicated operator depending on the time integrator chosen to solve the ODE. It has been shown that the discrete system is stable\footnote{It is stable as long as the Courant-Friedrich-Levy condition $\Delta t \leq \Delta x / c_h $ is fulfilled in hyperbolic systems, being $c_h$ the absolute value of the maximum eigenvalue}, consistent and convergent to the continuum solution if a locally stable time integrator, like Runge-Kutta (RK, see below) of at least $3^{\rm rd}$-order, is employed for the time evolution~\cite{Gustafsson:2013}.  

\subsection{Space derivative discrete operators}
\label{space_derivative_operators}

As it was mentioned before, ${\cal L}({\bf u})$ is an operator containing arbitrary spatial derivatives of the fields. This operator can be decomposed as
\begin{eqnarray}\label{PDEequationdecomposed}
    {\cal L}({\bf u}) = - \partial_k {\bf {\cal F}^k(u)} + {\bf {\cal S}(u,\partial u,\partial\partial u,...)} ~~.
\end{eqnarray}
where some of the terms, containing only first derivatives of the fluxes ${\bf {\cal F}^k(u)}$, are explicitly separated in order to take advantage of the existence of weak solutions in balance law equations. Notice that the fluxes ${\bf {\cal F}^k}$ might be non-linear but depend only on the fields, while that the generalized sources ${\bf {\cal S}}$ might depend not only on the fields but also on their spatial derivatives of arbitrary order.
This split allow us to define different discretization operators to deal with the fluxes and with the sources. In particular, finite difference schemes based on Taylor expansions, suitable for smooth solutions, will be applied to the generalized sources terms. However, the possible appearance of shocks in balance laws will require High-Resolution-Shock-Capturing (HRSC)
methods to discretize the fluxes\cite{Toro:1997}. We will therefore use a conservative scheme to discretize the fluxes and high-order spatial difference operators for the generalized sources. For instance, in two dimensions:

\begin{eqnarray}
\label{conservative_discretization}
L({\bf U}) = &-& \frac{1}{\Delta x} \left(F^x_{i+1/2,j} - F^x_{i-1/2,j}\right) \\
 &-& \frac{1}{\Delta y} \left(F^y_{i,j+1/2} - F^y_{i,j-1/2}\right)
 \nonumber \\
 &+& S(U, D\, U, ...)
 \nonumber 
\end{eqnarray}

Clearly, the crucial issue in the HRSC methods is how to compute the fluxes at the interfaces located at $x_{i\pm 1/2}$ such that no spurious oscillations appears in the solutions.

\subsubsection{Finite difference operators for smooth solutions}
\label{smooth_FD}

Suitable high-order discrete derivative operators for the generalized source terms $S(U,DU,...)$ can be found by using a Taylor expansion of the smooth solution around a specific position $x_i$ of the discrete grid. As a default, we will consider standard $4^{\rm th}$-order centered finite difference such that $D_i U \approx \partial_i u + {\cal O}({\Delta x}^5)$. In 2D, the $1^{\rm st}$-order derivative operators have the form
\begin{eqnarray}
  D_x U_{i,j} 
  &=& \frac{1}{12 \Delta x}
  \left( U_{i-2,j} - 8\, U_{i-1,j} + 8\, U_{i+1,j} - U_{i+2,j}       \right)   \nonumber \\
  D_y U_{i,j} &=& \frac{1}{12 \Delta y}
  \left( U_{i,j-2} - 8\, U_{i,j-1} + 8\, U_{i,j+1} - U_{i,j+2}       \right)  \nonumber
\end{eqnarray}

In some scenarios, like black hole evolutions within the Einstein equations that will be described later, it is extremely useful to use non-centered derivative operators to treat the advection terms of the equations, generically proportional to a vector $\beta^i$ (i.e., $\beta^i \partial_i u$)
Keeping the same fourth-order accuracy, one-side derivative schemes can be written as
\begin{eqnarray}
\beta^x \partial_x U_{i,j} = \frac{\beta^x}{12 \Delta x}
\begin{cases}
\biggl( - U_{i-3,j} +  6\, U_{i-2,j} - 18\, U_{i-1,j} \nonumber \\
+ 10\, U_{i,j} + 3 \,U_{i+1,j} \biggr)
~~,~~  \text{if}\ \beta^{x} < 0 \\    
\biggl( U_{i+3,j} -  6\, U_{i+2,j} + 18\, U_{i+1,j} \nonumber \\
- 10\, U_{i,j} - 3 \,U_{i-1,j}  \biggr) 
~~,~~  \text{if}\ \beta^{x} > 0 
\end{cases}
\end{eqnarray}

Second-order derivative operators can be constructed by applying twice the $1^{\rm st}$-order ones. This is a convenient
choice for the (commutative) cross-derivatives
\begin{eqnarray}
  D_{xy} U_{i,j}  = D_{yx} U_{i,j} &=& D_y \left( D_x U_{i,j} \right) ~~.
\end{eqnarray}
However, the stencil of the $2^{\rm nd}$-order derivative along a single coordinate direction (i.e., $xx$) would be twice larger than the one of the cross-derivatives. Therefore, with scalability in mind, it is preferable to change to a different $4^{\rm th}$-order operator which keeps the original stencil, namely
\begin{eqnarray}
  D_{xx} U_{i,j} = \frac{1}{12 \Delta x^2}
  \bigl( &-& U_{i-2,j} + 16 \, U_{i-1,j} -30 \, U_{i,j} \nonumber \\
    &+& 16\, U_{i+1,j} - U_{i+2,j}   \,\, \bigr) ~~.
\end{eqnarray}

Discrete numerical solutions might also contain unphysical high-frequency modes with a wavelength smaller than the grid size $\Delta x$ that can grow rapidly and spoil the real physical solution. These modes can be suppressed by including a small artificial Kreiss-Oliger (KO) dissipation along each coordinate direction~\cite{Calabrese:2004}. For instance, along the x-direction, the KO dissipation operator suitable for our $4^{\rm th}$-order scheme can be written as
\begin{eqnarray}
  Q^x_d\, U_{i,j} 
   &=& \frac{\sigma}{64 \Delta x}
  \bigl( U_{i-3,j}  -6 \, U_{i-2,j} + 15 \, U_{i-1,j}  \nonumber \\
   &&-20 \, U_{i,j} 
      + 15\, U_{i+1,j} - 6 \, U_{i+2,j} + U_{i+3,j}  \bigr) ~. \nonumber 
\end{eqnarray}
where $\sigma$ is a positive, adjustable parameter controlling the amount of dissipation added.

\subsubsection{Finite difference operators for non-smooth solutions}\label{nonsmooth_FD}

Equations which are intrinsically non-linear might develop shocks even from smooth initial data. The finite difference operators introduced previously rely on the high differentiability of the solutions, and therefore are not suited for these problems. HRSC methods are however designed to deal with possible shocks and discontinuities appearing in the solutions. The key point in HRSC methods is how to compute the fluxes at the interfaces located at $x_{i\pm 1/2}$. This calculation consists on two steps:

\begin{figure}[h]
	\includegraphics[width=7cm]{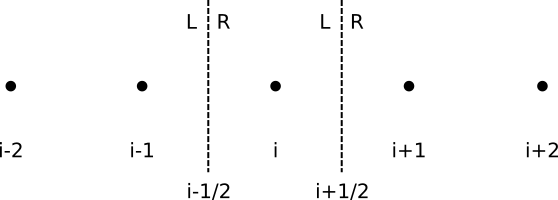}
	\caption{The computational uniform grid $x_i$. The left (L) and right (R) states reconstructed at the interfaces $x_{i\pm 1/2}$ are required to evolve the solution $U_i$. }
	\label{figure:weno}
\end{figure}

\begin{itemize}
  \item reconstruct the fields and fluxes, from the left (L) and from the right (R), in the interfaces between points.
  For instance, to evolve the field $U_i$ we will need to reconstruct the fields from left and right at neighboring interfaces $x_{i \pm 1/2}$, that is, $U^L_{i \pm 1/2}$ and $U^R_{i \pm 1/2}$ (see Fig.~\ref{figure:weno}). 
  \item  use a flux formula to compute the final flux at the interface, $F_{i \pm 1/2}$, that approximately solves the Riemann problem. One popular choice is the HLL flux formula~\cite{Harten:1983,Toro:1997}, which does not require the characteristic decomposition of the system
  \begin{eqnarray}
   F &=& \frac{1}{S^R - S^L} \left[ S^R F^L - S^L F^R \right. \\
   && ~~~~~~~~~~~~~  \left. S^R S^L (U^R - U^L)\right] \nonumber
  \end{eqnarray}
  
  where $F^L=F(U^L)$, $F^R=F(U^R)$ and $S^L,S^R$ are the fastest speed traveling to the left and to the right, respectively. They can be estimated as
  \begin{eqnarray}
     S^{L} &=& min ( {}^{(-)} \lambda^{L}, {}^{(-)} \lambda^{R}) ~,~ \\
     S^{R} &=& max ( {}^{(+)} \lambda^{L}, {}^{(+)} \lambda^{R}) ~.
  \end{eqnarray}
    
   A simplest and more robust choice assumes that $S^{L} = -S^{R} = S$. Substituting this expression into the HLL flux one can obtain the Local-Lax-Friedrichs (LLF) flux 
   \begin{equation}
   \label{LLF}
     F^{LLF} = \frac{1}{2} \left[ F^L + F^R
     - S (U^R - U^L)\right]
   \end{equation}
   that we will consider as the standard choice in our implementations.     
    
\end{itemize}

A important step on the discretization scheme is the reconstruction of the fields from the grid points $x_i$ into the interfaces located at $x_{i\pm 1/2}$. The reconstruction can be performed to the evolved fields, to the fluxes or to a combination of both. All these choices have advantages and disadvantages. For the tests presented later we apply the reconstruction to the evolved fields.

The reconstruction procedure can be performed at different orders. We have implemented several of the most commonly used reconstructions, like PPM~\cite{Colella:1984}  and MP5~\cite{Suresh:1997}, and other implementations like the FDOC families~\cite{Bona:2009} which are almost as fast as centered Finite Difference at the cost of some bounded oscillations near the shock region. Here we present a short summary of the Weighted-Essentially-Non-Oscillatory (WENO) reconstructions~\cite{Jiang:1996,Shu:1998}, which is our preferred choice for their flexibility (i.e., they can achieve any order of accuracy) and robustness. The detailed implementation of the WENO flavors used here can be found in Appendices~\ref{appW3} and \ref{appW5}, while that details of the other methods can be found in a recent review~\cite{Balsara:2017}. An upwind biased $(2k-1)^{\rm th}$-order approximations to the function $u(x)$ at the neighbor interfaces around the cell $U_i$, denoted by ${}^R U_{i-1/2}$ and ${}^L U_{i+1/2}$, can be obtained in the following way:
\begin{itemize}
  \item Obtain the k reconstructed values ${}^L U^{(r)}_{i+1/2}$
  and ${}^R U^{(r)}_{i-1/2}$ of $k^{\rm th}$-order accuracy,
  \begin{eqnarray}
      ^L U^{(r)}_{i+1/2} &=& \sum_{j=0}^{k-1} c_{r,j} U_{i-r+j} 
      \nonumber \\
      ^R U^{(r)}_{i-1/2} &=& \sum_{j=0}^{k-1} c_{r-1,j} U_{i-r+j}       
  \end{eqnarray}
  with $r=0..k-1$. For instance, the coefficients $c_{r,j}$ for the cases $k=2$ and $k=3$ can be found in Table~\ref{coefficients_crj}.
  
  \begin{table}[ht]
  \caption{Coefficients $c_{r,j}$ for $k=2$ (left) and $k=3$ (right).}
  \begin{minipage}[b]{0.4\linewidth}
  \centering
  \begin{tabular} {c c c c c}
   r     & \vline & j=0 & \vline &  j=1   \\
  \hline 
   -1   & \vline & 3/2 & \vline & -1/2    \\
   0   & \vline & 1/2 & \vline &  1/2     \\
   1     & \vline & -1/2 & \vline  & 3/2   \\
  \end{tabular}
  \end{minipage}
  \hfill
  \begin{minipage}[b]{0.5\linewidth}
  \centering
  \begin{tabular} {c c c c c c c}
   r     & \vline & j=0 & \vline &  j=1 & \vline &  j=2  \\
  \hline 
  -1   & \vline & 11/6 & \vline & -7/6 & \vline &  1/3 \\
   0   & \vline & 1/3  & \vline &  5/6 & \vline & -1/6 \\
   1   & \vline & -1/6 & \vline &  5/6 & \vline &  1/3 \\
   2   & \vline & 1/3  & \vline & -7/6 & \vline & 11/6 \\
  \end{tabular}
  \end{minipage}
  \label{coefficients_crj}
  \end{table}
    
  \item Find the smooth indicators ${}^L \beta^{(r)}_{i+1/2}$ and ${}^R \beta^{(r)}_{i+1/2}$, that will depend on the order $k$
  
  \item Find the $(2k-1)^{\rm th}$-order reconstruction 
  \begin{eqnarray}
      U^{L}_{i+1/2} &=& \sum_{r=0}^{k-1} \omega^{(r)}_{i+1/2} {}^L U^{(r)}_{i+1/2} 
      \nonumber \\
      U^{R}_{i-1/2} &=& \sum_{r=0}^{k-1} {\tilde \omega}^{(r)}_{i-1/2} {}^R U^{(r)}_{i-1/2} 
  \end{eqnarray}
  with $r=0..k-1$. The weights, $\omega^{(r)}_{i+1/2}$ for the left and   ${\tilde \omega}^{(r)}_{i+1/2}$ for the right, can be constructed in the following way:
  \begin{eqnarray}
      \omega^{(r)}_{i+1/2} &=& \frac{\alpha^{(r)}_{i+1/2}}{\sum_{s=0}^{k-1}\alpha^{(s)}_{i+1/2}} ~,~
      \alpha^{(r)}_{i+1/2} = \frac{d_r}{(\epsilon + {}^L \beta^{(r)}_{i+1/2} )^2}
      \nonumber \\
      {\tilde \omega}^{(r)}_{i-1/2} &=& \frac{{\tilde \alpha}^{(r)}_{i-1/2}}{\sum_{s=0}^{k-1} {\tilde \alpha}^{(s)}_{i-1/2}} ~,~
      {\tilde \alpha}^{(r)}_{i-1/2} = \frac{{\tilde d}_r}{(\epsilon + {}^R \beta^{(r)}_{i-1/2} )^2} \nonumber
      \end{eqnarray}  
  where ${\tilde d}_r = d_{k-1-r}$ and $\epsilon$ is a very small number to avoid division by zero. The coefficients $d_r$ for the cases $k=2$ and $k=3$ are
  \begin{eqnarray}
      k&=&2 ~~~~~~  d_0 = 2/3 ~,~d_1 = 1/3  \\
      k&=&3 ~~~~~~  d_0 = 3/10 ~,~d_1 = 6/10 ~,~d_2= 1/10 ~~.\nonumber
  \end{eqnarray}

\end{itemize}

\subsection{Runge-Kutta time integrator}

Locally stable time integrators ensure the stability and convergence of the solution of the evolution system. RK schemes of at least $3^{\rm rd}$-order are locally stable and are relatively easy to implement for solving the ODEs associated to the semi-discrete system. A RK scheme with $s$ stages, applied to the system (\ref{semidiscrete}) without dissipation, allows to express the solution at the next time-step $U^{n+1}$ as a combination of several auxiliary intermediate steps $U^{(i)}$~\cite{Butcher:2008}, namely 
\begin{eqnarray}\label{eRK}
   {U}^{(i)} = {U}^n &+& \sum_{j=1}^{i} {b}_{ij} {k}_j  ~~,~~
   k_j = \Delta t \, L(U^{(j)})
\nonumber \\
  {U}^{n+1} = {U}^n &+& \sum_{i=1}^{s} {c}_{i} k_i
\nonumber
\end{eqnarray}
The matrices $B= (b_{ij})$ have dimension $s \times s$ and lead to a scheme of order $p$ (i.e., the error is ${\cal O}(\Delta t^{p+1})$). For simplicity, explicit schemes (i.e., such that ${b}_{ij} = 0$ for $j \geq i$) are preferred over more complicated choices. A RK scheme is characterized by this matrix and the coefficient vector $c_i$, which can be represented by a tableau in the usual Butcher notation~\citep{Butcher:2008}:
\begin{table}[h!]
\label{butcher_tableau}
\begin{minipage}{1in}
\begin{tabular} {c c c}
${a}$  & \vline & ${B}$  \\
\hline 
              & \vline & ${c}^T$  \\
\end{tabular}
\end{minipage}
\end{table}

where the coefficients $a_i$ 
used for the treatment of non-autonomous systems are given by the consistency relation $ {a}_{i} = \sum_{j=1}^{i-1}~ {b}_{ij}$. These schemes can be denoted as RK$(s,p)$,
where the doblet $(s,p)$ characterizes the number of $s$-stages of the explicit scheme and the order $p$ of the scheme. It is possible to construct RK schemes of order $p=s$ up to $p \le 4$, making this choice optimal. A very well known $4^{\rm th}$-order RK 
which remains stable under quite large time-steps is given in Table~\ref{RK44-SSP3}. 

\begin{table}[h]
\label{RK44-SSP3}
\caption{Tableau for a very common explicit RK(4,4).}
\begin{minipage}{3.4in}
\begin{tabular} {c c c c c c}
 0     & \vline & 0  &  0  &  0  & 0  \\
 1/2   & \vline & 1/2  &  0  &  0  & 0  \\
 1/2   & \vline & 0  &  1/2  &  0  & 0 \\
 1     & \vline & 0  & 0 & 1 & 0 \\
\hline 
   & \vline &  1/6 & 2/6 & 2/6 & 1/6 \\
\end{tabular}
\end{minipage}
\end{table}
Therefore, the explicit implementation of the RK(4,4) is just:
\begin{eqnarray}\label{RK4_implementation2}
   {U}^{(1)} &=& {U}^n \\
   {U}^{(2)} &=& {U}^n + \frac{1}{2} k_{1} \\
   {U}^{(3)} &=& {U}^n + \frac{1}{2} k_{2} \\
   {U}^{(4)} &=& {U}^n + k_{3} \\
   {U}^{n+1} &=& {U}^n + \frac{1}{6} 
   \left( k_1 + 2 k_2  + 2 k_3 + k_4  \right)  
\end{eqnarray} 
where $k_i = \Delta t \, L({U}^{(i)})$.


\section{The AMR algorithm}

One way to use efficiently the computational resources is increasing the grid resolution only on the localized regions of the simulation domain where the dynamics is more demanding and higher resolution is required to improve the accuracy of the solution. A mature and well-established strategy is the AMR, which introduces new additional grid levels with higher resolution on specific regions which might change dynamically with the solution.
The AMR algorithm specifies how the solution on multi-processor and multi-levels is evolved, and in particular, how the information on the different domain boundaries is shared among the multiple processors. In our approach, a generic AMR algorithm is constructed by using the basic blocks (i.e., routines and functions) provided by SAMRAI. The algorithm skeleton for a problem with $L$ refinement levels, for the specific case of RK integrators with $S$ sub-steps, could be written as follows:

\noindent\fbox{%
	\begin{minipage}{\dimexpr\linewidth-2\fboxsep-2\fboxrule \relax}
\begin{algorithmic}
{\tt
 \State initialization
 \State refinement tagging
 \While{not simulation end}
   \ForAll{level l= 0,L}
     \ForAll{RK sub-steps s= 1,S}
	   \State calculate rhs 
	   \State integrate time
	   \If{last Runge Kutta substep}
	   	   \State restrict from l to l-1
		   \State level synchronization(l - 1)
	   \EndIf	   
   	   \State level synchronization(l)
   	   \State prolong from l-1 to l 
   	   \State calculate physical boundaries
	 \EndFor
	 \If{has to regrid(l)}
	   \State refinement tagging(l)
	 \EndIf
   \EndFor
 \EndWhile }
\end{algorithmic}
\end{minipage}%
}

The algorithm calls \textbf{refinement criteria} to decide which regions need additional levels with smaller grid sizes to obtain an accurate solution. Once the solution is defined in all levels  the simulation can start. The procedure to integrate a time-step is repeated over and over until reaching the final simulation time. The fields must be evolved in all grids each timestep, starting from the coarsest level $l=0$ to the finest one $l=L$. Each time integration is performed by using a RK with S sub-steps. Therefore, the intermediate auxiliary states $U^{(i)}$ and the final one $U^{n+1}$ must be computed at each level. The right-hand-side of the evolution equations, which involves spatial derivatives, need to be computed at each of these sub-steps, by using the discrete spatial operators described in the previous section. Notice also that the nearby zones outside the boundary of the fine levels must be filled with points of the same resolution in order to accurately evolve the solution. This procedure is called \textbf{prolongation} and it usually involves interpolation from the coarse grid level into the fine one. After computing each intermediate RK-step the fields need to be synchronized among the different processors on level $l$ in order to fill the boundaries of the domains splatted in each processor with the correct updated data. Similarly, after finishing all the steps of the RK, we need to inject the solution of the fine level $l$ into the coarse one $l-1$, a procedure known as \textbf{restriction}. After the values on the coarse grids have been updated, the information on the level $l-1$ must be again synchronized among processors.

\subsection{Refinement criteria, restriction and prolongation}

There are several strategies to decide which regions need
more resolution to be accurately resolved by including additional grid levels with higher resolution. These strategies rely
on going through the points of the coarsest level and evaluating some refinement criteria, such that an additional level can afterwards be added in the tagged regions. This process can be repeated in the new refined levels until some condition is fulfilled, either on the refinement criteria or on a maximum allowed number of levels. There are two refinement tagging strategies provided by SAMRAI integrated in Simflowny.
\begin{itemize}
\item \textbf{Fixed Mesh Refinement (FMR).} The user specifies statically a set of boxes where the refinement is located. Every level allows different boxes as long as they are nested in coarser level boxes.
\item \textbf{Adaptive Mesh Refinement (AMR).} The user sets a criteria (i.e., a measurement of the error or a function of the fields surpassing certain threshold) used to dynamically calculate the cells to be refined. 
\end{itemize}

Notice that fixed and dynamical tagging strategies (i.e., FMR and AMR) can be combined in the same simulation. As the simulation evolves, the AMR tagging criteria will likely change, implying that new regions will be refined and old ones will be disposed of. This re-meshing procedure is performed periodically. 

If a new refinement level is added dynamically during the simulation (i.e., or the region of a given level increases due to the dynamical AMR criteria), the domain of that grid increases with respect to the coarser level. The new grid points on the fine level are set by the prolongation procedure,
interpolating the solution from the coarse grid into the fine one. This spatial interpolation must be more accurate than the spatial derivative operators in order to prevent the spoiling of the scheme accuracy. One of the simplest and most efficient options is to use Lagrange interpolating functions. Given a solution $U_i$ at the position $x_i$, one can construct a Lagrangian polynomial function of order $k$ passing through a $k+1$ set of points $\{(x_1,U_1), (x_2,U_2),...(x_k,U_k),(x_{k+1},U_{k+1})\}$, namely
\begin{equation}
  p(x) = \sum\limits_{j=1}^{k+1} U_j l_j(x) ~~~,~~~
  l_j(x) = \prod_{\substack{m=1 \\ m \neq j}}^{k+1} \frac{x - x_m}{x_j - x_m}
\end{equation}
where $x$ is the point position in which the value is interpolated. To construct a symmetric Lagrangian polynomial of $5^{\rm th}$-order, suitable for our $4^{\rm th}$-order spatial scheme, six points are required (i.e, three at each side of the point to be interpolated). Such Lagrangian polynomial interpolation can be simplified for the centered point $x=x_0$, namely
\begin{eqnarray}
  p(x_0) &=& \frac{1}{256}  \biggl[ 150 (U_{x-1} + U_{x+1}) 
                 - 25 (U_{x-2} + U_{x+2}) \nonumber \\
           &&  ~~~~~  + 3 (U_{x-3}+U_{x+3}) \biggr]
\end{eqnarray}
In structured grids it is common to choose refined grids such that the points of the coarse grid also exist in the fine grid (i.e., the ratio between their resolutions is $2^p$), so this interpolation is the only one required.

Since we are interested on MHD problems involving non-smooth solutions, it is relevant to study if this interpolation is suitable when shocks and discontinuities are present in our simulation.
Indeed, this simple Lagrange interpolation has been compared to a WENO interpolation for systems of equations with non-smooth solutions~\cite{Sebastian:2003}. The comparison indicates that the simple and efficient Lagrange interpolation, combined with a WENO finite difference method to discretize the derivatives during the evolution, suffices for the domain interface treatment to retain high-order of accuracy and essentially non-oscillatory properties even for strong shocks~\cite{Sebastian:2003}.

The restriction procedure is complementary to the prolongation. In the restriction,  on the regions with overlapping grids, the data from a fine level is injected into a coarse one. If the points of the coarse grid also exist in the fine grid (i.e., like when the ratio between the two resolutions is $2^p$), the restriction is quite straightforward and only implies copying directly data from the fine level to the coarse one.

\subsection {Sub-cycling in time}

A necessary condition for the stability of  explicit numerical schemes of hyperbolic systems is that the time step must satisfy the CFL condition $\Delta t \le \lambda_{\rm CFL} \Delta x$, with $\lambda_{\rm CFL}$ a factor depending on the dimensionality of the problem and the specific time integrator. When there are multi-levels $l=0..L$, the solution on each refinement level can be evolved in a stable way by using the time-step corresponding to the finest grid resolution ${\Delta x}_{L}$, ensuring that all the grids satisfy the CFL condition. This is however a very inefficient choice, since coarser grids are evolved with a time-step much smaller than the one allowed by their local CFL condition.

A common way to avoid such a restriction is by evolving the solution with sub-cycling in time, meaning that each grid uses the largest $\Delta t$ as set by its local CFL condition. This means that the finer grids must perform two or more time-steps for each one of the coarse grid. In this case, it is not clear how to evolve the interior points of the fine grid at the refinement boundary, since the solution is not evaluated at the same time on the coarser grid. There have been several well motivated strategies to fill in this missing information:
\begin{itemize}
 \item {\bf Tappering}. The fine grid is extended by a number of points given by $N_{\rm ext} = f_{\rm res} N_{\rm st} N_{\rm RK}$, for a resolution ratio of $f_{\rm res}$, stencil points $N_{\rm st}$ and RK time sub-steps $N_{\rm RK}$, on each direction perpendicular to the refinement boundary. This way, points at the boundary can be evolved without any intermediate prolongation~\cite{Lehner:2006}. The boundary points at the end of the time step of the fine grid (i.e., when it reaches $U^{n+1}$) are inside the numerical domain of dependence of the extended initial fine grid. This algorithm is computationally expensive and it is difficult to achieve a good scalability because involves extending each refinement grid by a large number of points in each direction. For instance, with a $4^{\rm th}$-order RK and $4^{\rm th}$-order space discretization it would be around 16 points on each side of the fine grid. However, it is very accurate, since it minimizes boundary reflections at the  interfaces between levels.
 
 \item {\bf Berger-Oliger algorithm (BO1)}. The solution of the coarser grid is evolved first up to $n+1$. Then, with the information from $\{U^{n+1},U^{n}\}$, we can interpolate in time to calculate the solution at the required times of the RK schema of the finer grids.  Spatial interpolation (prolongation) is also required to fill the points in the positions needed by the spatial discretization scheme. This algorithm is cheap, fast and efficient, since it requires to interpolate only in a number of points equal to the stencil of the spatial discretization scheme. For instance, it just requires 3 points in the ghost-zone for a $4^{\rm th}$-order centered derivatives with $6^{\rm th}$-order dissipation. The drawback of this simplest original version is that it is only $1^{\rm st}$-order accurate in time.

 \item {\bf Berger-Oliger with dense output interpolator (BO)}. The  original Berger-Oliger algorithm can be improved by using additional information to increase the accuracy of the interpolation scheme by either (i) including other time-levels $\{U^{n-1},U^{n-2},...\}$, or (ii) including the intermediate RK solutions $U^{(i)}$. This last option, that we will consider here, is commonly known as  dense output interpolator, and its implementation for some RK schemes is discussed in detail in Appendix~\ref{appDense}.

 \item {\bf Berger-Oliger without order reduction (BOR)}. The first step of the algorithm is similar to the BO one, using information from all the sub-steps of the RK (i.e., $\{U^n,U^{(i)},U^{n+1}\}$) to build an internal dense output interpolator of order $q=p-1$. However, in the second step this interpolator is used for computing all the time derivatives of the fine grid~\cite{McCorquodale:2011,Mongwane:2015}. By using the standard RK formula with these time derivatives it is possible to calculate the solution at each RK sub-step and achieve a final scheme at least order $q$ in time. This algorithm, which is discussed thoroughly in  Appendix~\ref{App_BOR}, is fast, efficient and very accurate. Moreover, we have extended the algorithm to allow arbitrary resolution ratios between consecutive AMR grids.

\end{itemize}

The prolongation, restriction and number of executions depend on whether sub-cycling is active and which option from the previous ones is being considered. Currently, there are four available AMR time integrations in Simflowny: no sub-cycling in time, tappering, standard BO and BOR. The latter will be our preferred choice.


\section{Tests}

We will focus on two specific simple models to test the implementation of our mesh-refinement algorithms and our numerical schemes. First, we  consider the scalar wave equation, that will allow us to check the accuracy and convergence properties of the discrete spatial derivative operators for smooth solutions. More important, it provides a very controlled setup to test the different strategies of sub-cycling in time.
The second model is the Newtonian MHD equations, that will allow us to check the numerical schemes for non-smooth solutions. We also check that our preferred choices for the AMR/FMR algorithms do still work well for these kind of systems. Although the space discrete derivatives are calculated with different operators for these two models, the integration in time of the semi-discrete system of ODEs is performed in both cases by using a $4^{\rm th}$-order RK.

One of the most important analysis quantities to validate the numerical schemes in our tests is the convergence factor. Let us consider the numerical solutions of a scalar or a vector component discretized field $U_L$ obtained with three different resolutions $\Delta x_L$, such that $r=\Delta x_0/\Delta x_1=\Delta x_1/\Delta x_2 > 1$ (typically, $r=2$). The convergence order $n_c$ of the numerical solution at a given timestep $t$ can be defined as
\begin{equation}
  n_c(t) = \log_r \left(\frac{||U_0 - U_1||_1}{||U_1 - U_2||_1}\right)
\end{equation}
where $||U_m - U_n||_1 = \Sigma_{\vec{i}} |U_m^{\vec{i}} - U_n^{\vec{i}}|$ is the L1-norm of the difference of the two discretized fields, and the sum is performed over all the set of indexes $i_k$, k=1...N, identifying the N-dimensional grid with the lowest resolution. In our tests, we will compare the numerical convergence order with the nominal one expected for every scheme. In the case in which we can compare the numerical solution $U$ to an analytical or reference solution $U_{\rm ref}$ (e.g., a high-resolution run), we evaluate also the relative error as:
\begin{equation}\label{def:relative_error}
  \epsilon_{\rm num}(t,N) = \frac{||U - U_{\rm ref}||_1}{||U_{\rm ref}||_1}
\end{equation}
which will depend on the number of points $N$ employed. For a given time $t$, $\epsilon_{\rm num}(N)$ is related to the convergence order by $d\epsilon_{\rm num}/dN = - n_c$.

\subsection{Wave equation}

The simple wave equation may be written as a system with partial derivatives of $1^{\rm st}$-order in time and $2^{\rm nd}$-order in space, namely
\begin{eqnarray}
  \partial_t \phi &=& - \Pi \\
  \partial_t \Pi &=& -{\eta}^{ij} \partial_i \partial_j \phi 
\end{eqnarray}
where $\eta^{ij}=1$ for $i=j$ and zero otherwise. The simulations are performed in a two-dimensional narrow channel, although the choice of the initial 
data (i.e,. a Gaussian in the x-direction) restricts the problem to be one-dimensional. The domain along the non-trivial direction is $x \in [-2, 8]$ with periodic boundary conditions. We choose an initial configuration given by a time-symmetric pulse centered at $x=0$, namely
\begin{eqnarray}
  \phi_0(x) = \phi (x,t=0) = e^{-x^2/{\varrho^2}} ~~~,~~~
  \Pi = 0 ~~~~,
\end{eqnarray}
with $\varrho=0.173$. As time evolves, the initial Gaussian profile splits in two identical pulses propagating in opposite directions. These two pulses overlap again at the initial location after a full crossing time $t_{\rm CT}=10$. We evolve this problem with $4^{\rm th}$-order space differencing and $6^{\rm th}$-order KO dissipation, such that the semi-discrete problem is consistent to the continuum one to $4^{\rm th}$-order accuracy in $\Delta x$. With a single grid level it is straightforward to show that our numerical solution converges to the analytical one with the expected $4^{\rm th}$-order.

\begin{figure}[h]
 \includegraphics[width=8cm]{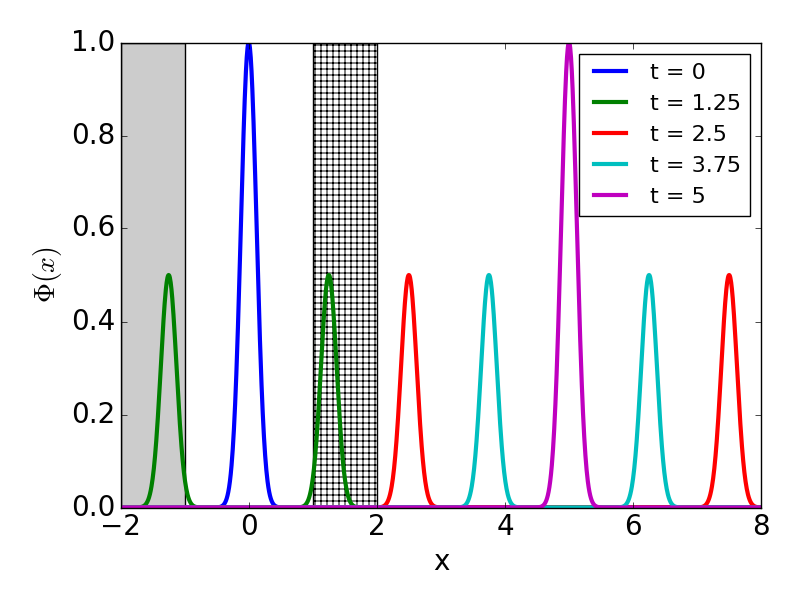}
 \caption{{\em Wave equation with FMR}. Scalar field at different times of the evolution. The pulse traveling to the right crosses the refined region located in the interval $x\in[1,2]$ (marked with a mesh). The scalar field norm is integrated in the interval $x\in [-2,-1]$ (grey shade) to get an estimate on the reflections.} 
 \label{figure:solution}
\end{figure}

The problem becomes more interesting by including an additional fixed grid level for $x\in[1,2]$, with twice the resolution of the coarse original grid, which is set to either $\Delta x=\{ 1/40, 1/80, 1/160\}$. The pulse traveling to the right will cross the refined region and then interact with the one traveling to the left before returning to its initial position. The pulse traveling to the left will also cross the  refined region after the interaction. The solution at different times is displayed in Figure~\ref{figure:solution}, together with the grid level with finer resolution (i.e., the squared region).

\begin{figure}[h]
 \includegraphics[width=8cm]{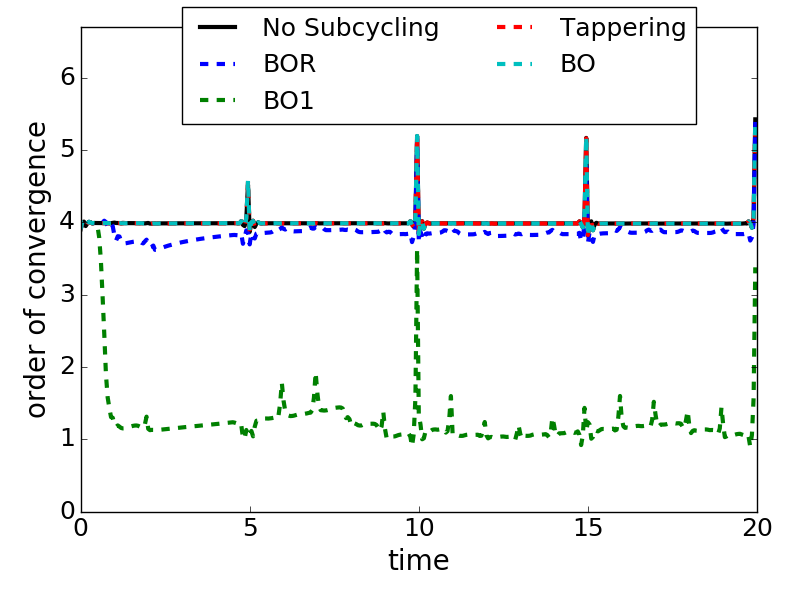}
 \caption{{\em Wave equation with FMR}. Convergence order of the solution with different mesh-refinement strategies, using $\Delta x=\{ 1/40, 1/80, 1/160\}$. 
  For this simple problem, $4^{\rm th}$-order convergence is achieved with all the time sub-cycling strategies except for the linear BO1. }
 \label{figure:reflections_rk4}
\end{figure}

The convergence of the numerical scheme will depend now not only on the order of the space discretization $p$ and time integrator $q$, but also on the choice of time refinement algorithm. We have considered here five different choices; no sub-cycling, tappering, standard linear Berger-Oliger (BO1), Berger-Oliger with a dense output interpolator of order $q-1$ (BO) and Berger-Oliger without order reduction (BOR). The convergence rate for these time sub-cycling strategies is displayed in Figure~\ref{figure:reflections_rk4}, showing that all of them, except for the linear BO1, achieve the expected $4^{\rm th}$-order. From now on, we can discard BO1 completely and focus only on the other cases.

Another important feature that can strongly depend on the sub-cycling in time strategy is the presence of reflections due to the change of dispersion/group numerical velocities when a pulse crosses a refined grid. This effect is also present in our test problem since both pulses, traveling to the left and right, cross the refined grid every turn. An easy and straightforward way to estimate these reflections is by computing the integral of the scalar field norm in the interval $x \in [-2,-1]$, namely
\begin{equation}
   Q \equiv \int\limits_{-2}^{-1} |\phi| dx
\end{equation}

 This quantity Q will first measure the (integral of the) pulse traveling directly to the left, and then the reflections of the pulse traveling to the right as it crosses the fine region (i.e., one as it goes from coarse to fine at $x=1$ and another as goes back from fine to coarse at $x=2$). 

\begin{figure}[h]
 \includegraphics[width=8.0cm]{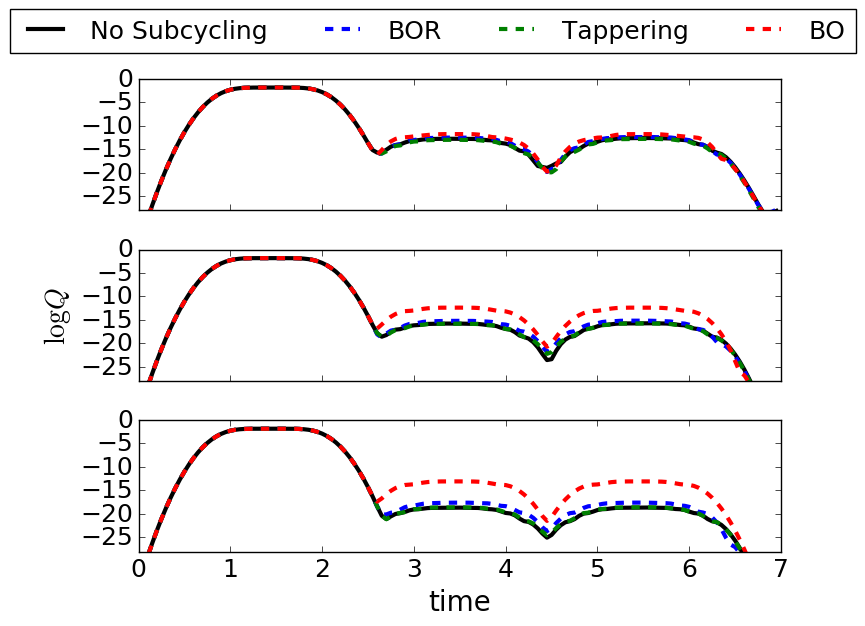}  
 \caption{{\em Wave equation with FMR}. Integral in the spatial interval $x\in[-2,-1]$ of the norm of the scalar field for three different resolutions $\Delta x=\{ 1/40, 1/80, 1/160\}$ (from the lowest resolution on the top to the highest at the bottom) of the coarse grid. The first bump is produced by the pulse traveling to the left,  which is a specular image of the  one traveling to the right. The (much smaller) second and third bumps correspond, respectively, to the reflections produced by the pulse traveling to the right as it enters and exits the refined region. A calculation of the convergence order confirms that the solution obtained by using BO converges only to $1^{\rm st}$-order, while that the other three cases converge to $4^{\rm th}$-order.}
 \label{figure:reflection1600}
\end{figure}

Figure~\ref{figure:reflection1600} displays this integral as a function of time, showing the three stages (i.e., pulse traveling to the left, followed by the reflections as it enters on and exits from the fine region). 
A lower bound for these reflections is given by the case without sub-cycling, since the solution on different grid levels is evolved with the same $\Delta t$ and the internal boundary conditions are just interpolated between solutions at the same time level. The expectation is that the Tappering strategy should be very close to the no sub-cycling one, since the grids have been extended such that the points at the refinement boundary can be evolved a full time-step without communicating information between grid levels. The standard BO algorithm involves communication and time interpolation between refinement grids. Nevertheless, since these additional calculations are restricted only to a few points near the interface boundary, it is still much more efficient than tappering. Despite  converging globally with the same accuracy than the RK integrator, the BO displays the largest reflections. The BOR strategy improves these results, decreasing the reflections almost to the level of the Tappering and no sub-cycling cases. These results can be understood easily by computing the convergence factor of this integral, which shows that all the cases converge to $4^{\rm th}$-order, except the BO which converges only to $1^{\rm st}$-order.
This means that, although the global solution, which is dominated by the main pulse, converges with the expected order to the continuum one, the  reflections do not converge with the same rate when using BO: even when using a high-order time dense output interpolation, the errors of the different time sub-steps do not cancel out automatically and convergence is spoiled. 
These results indicate that BOR is the most efficient time refinement strategy, since it does not require much memory overhead (i.e., needs a small ghost zone), it is quite fast and introduces very little reflections, at the same level than Tappering and no sub-cycling.

\begin{figure}[h]
	\includegraphics[width=8.6cm]{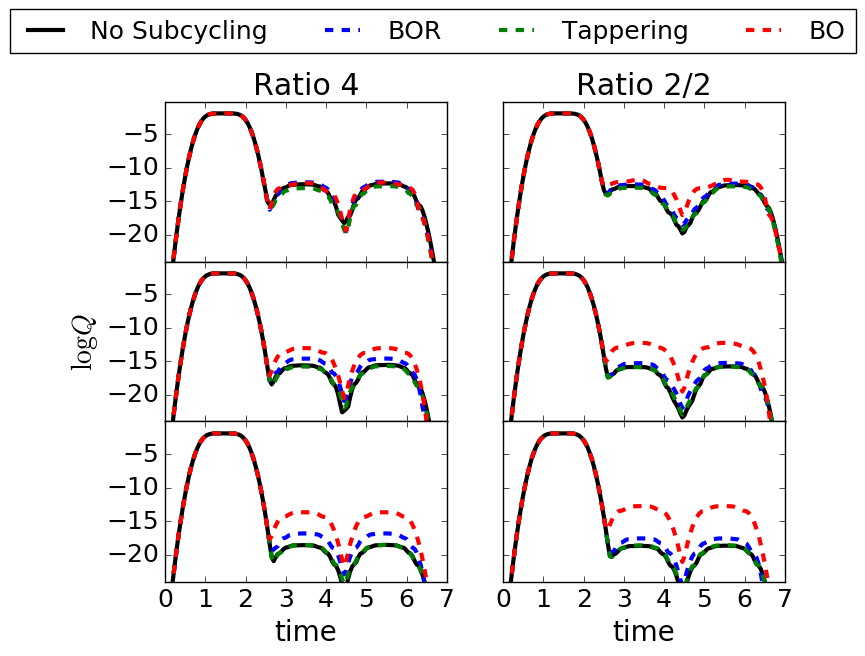}
 \caption{{\em Wave equation with FMR}. Same as Figure~ \ref{figure:reflection1600} but with either one refinement level with ratio $4$ or two levels with ratio $2$. The BOR algorithm reduces considerably the reflections in both grid setups with respect to BO, reaching almost the level achieved without sub-cycling.
 }
 	\label{figure:reflectionRatio4}
\end{figure}

As we will show later when evolving black holes with the Einstein equations, the scaling properties of the problem might change
depending on the grid structure, and on particular, on the memory load of the finest
grid level. For this reason it is important to test also grid configurations
with resolution ratios larger than 2. In Figure~\ref{figure:reflectionRatio4} the pulse reflections are plotted for two different grid setups similar to the previous one, but with either two levels of refinement (instead of one) or with a single level with a ratio 4 between grid resolutions.
The results with ratio 2/2 and with ratio 4 are quantitatively similar, with slightly more reflections in the latter case.

\begin{figure}[h]
	\includegraphics[width=8cm]{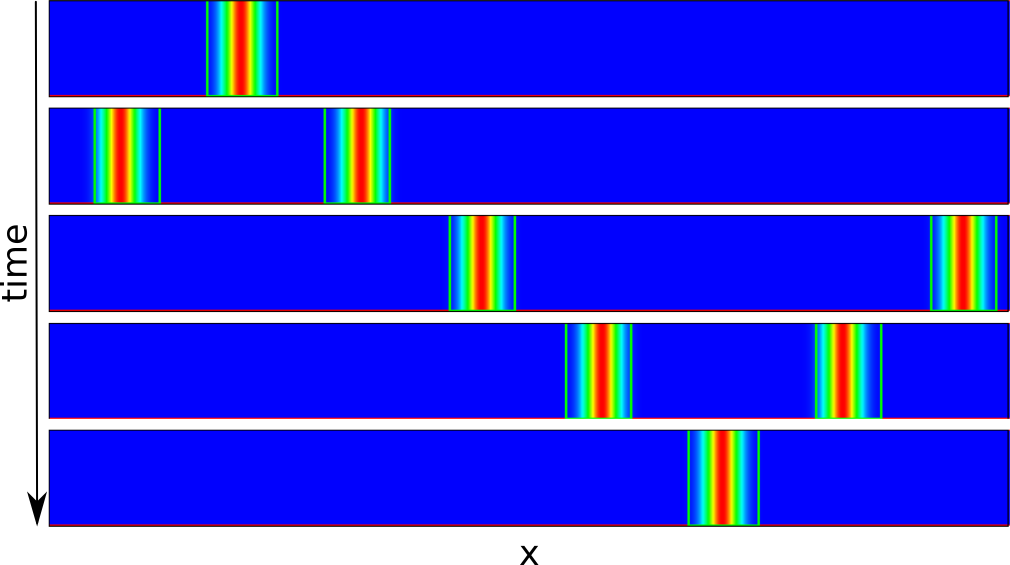}
 \caption{{\em Wave equation with AMR}. Scalar field and the refinement
 box (marked with green lines) at different times, from top to bottom, of the evolution covering half a crossing time $t=\{0, 1.25, 2.5, 3.75, 5\}$. The simulation is performed in a two-dimensional narrow channel, although the choice of the initial 
 data (i.e,. a Gaussian in the x-direction) restricts the problem to be one-dimensional.}
	\label{figure:solution_wave_AMR}
\end{figure}

\begin{figure}[h]
 \includegraphics[width=8cm]{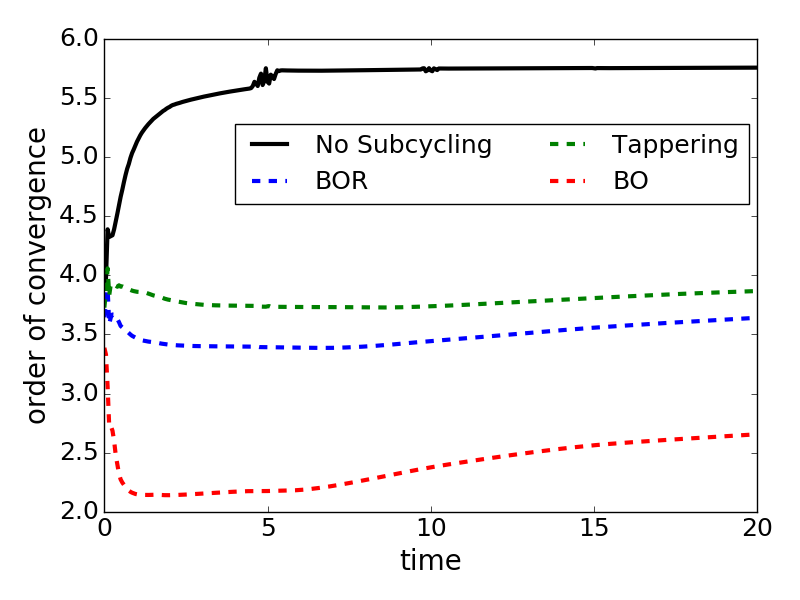}
\caption{{\em Wave equation with AMR}. Convergence rate using $\Delta x=\{ 1/40, 1/80, 1/160\}$. Notice that BO fails again to achieve the expected $4^{\rm th}$-order convergence rate due to the interaction of the pulse with the moving refinement interface. The case without sub-cycling shows super-convergence. }
 \label{figure:amr}
\end{figure}

Finally, we can study the the behavior of the solution by using the different time-refinement algorithms in a problem with full AMR, by setting a refinement criteria such that the refined grid follows the pulses (i.e., the grid is refined whenever $\phi \geq 10^{-3}$). The solution and the refined region at different times are displayed in Figure~\ref{figure:solution_wave_AMR}, and the corresponding convergence factor in Figure~\ref{figure:amr}. There are several interesting features of these results.
The BO algorithm converges globally to a $2.5$ order, but both the Tappering and the BOR converges to almost $4^{\rm th}$-order, the factor expected for RK4. The case without sub-cycling shows a super-convergence with a factor close to 6.

The conclusions of these tests with the wave equation can be summarized as follow: (i) our implementation of the RK and the discrete spatial operators for smooth solutions are correct, (ii) the BOR sub-cycling strategy for FMR/AMR is the most accurate and maintains the convergence rate of the numerical scheme, and (iii) smooth solutions with resolution ratios larger than 2 between consecutive grid levels are equivalent on accuracy to those with ratio 2.

\subsection{Newtonian ideal MHD equations}

The ideal MHD equations, describing a magnetized perfect
fluid, can be written in terms of the total energy $E$ and the momentum density $S_i$,
\begin{equation}
   E=\frac{1}{2}\rho v^2 + \rho \epsilon + \frac{B^2}{2} ~~~,~~~ S_i =  \rho v_i
\end{equation}
where $\rho$ is the fluid density, $\epsilon$ its specific internal energy, $v_i$ its velocity and $B^i$ the magnetic field.
Within these definitions, the complete set
of evolution equations can be written as
\begin{eqnarray}
   \partial_t \rho &+& \partial_k [\rho v^k] = 0 \\
   \partial_t E &+& \partial_k \left[ \left( E + p + \frac{B^2}{2}\right) v^k 
             -  (v_j B^j) B^k \right] = 0 \\
   \partial_t S_i  &+&
   \partial_k \left[ \rho v^k v_i + \delta^k_i \left(p+\frac{B^2}{2} \right) 
              - B^k B_i \right] = 0 \\
  \partial_t B^{i} &+& 
  \partial_k [v^k B^i - v^i B^k + \delta^{ki} \psi ]  =   0 \\
  \partial_t \psi &+& c_h^2 \partial_i B^i = 
     - \kappa_{\psi} \psi
\end{eqnarray}
where $\psi$ is a scalar introduced to enforce dynamically the solenoid constraint $\nabla_i B^i$. This divergence cleaning
approach allows to propagate the constraint violations with
a speed $c_h$ and damp them exponentially in a timescale $1/\kappa_{\psi}$~\cite{2002JCoPh.175..645D}. 

In order to close this system of equations one needs to provide the Equation of State (EoS) relating the pressure to the other fluid variables, $p=p(\rho,\epsilon)$. A good approximation,  that allows to recover the fluid variables from the evolved fields through algebraical relations, is to consider the ideal gas EoS $p = (\Gamma-1) \rho \epsilon$, where $\Gamma$ is the adiabatic index.

\subsubsection{Circularly polarized Alfven wave}

Our first benchmark test of the numerical scheme for fluids is the 2D circularly polarized Alfven wave problem~\citep{Toth:2000}, which is the advection of a smooth solution of the ideal compressible MHD equation in a periodic 2D plane. The initial conditions for the components parallel ($\parallel$) and perpendicular ($\perp$) to the wave-vector $\bf{k}$ 
are set as follows:
\begin{eqnarray}
 && \rho =1 ~~,~~ p=0.1 ~~,~~ B_\parallel = 1 ~~,\\
 && B_\perp  = v_\perp = 0.1\sin (2\pi x_\parallel) \\
 && B_z  = v_z = 0.1\cos (2\pi x_\parallel) 
\end{eqnarray}
where $\vec{k}$ is contained in the ${x,y}$ plane, and $x_\parallel = \vec{k}\cdot\vec{x} = (x\cos\alpha_k + y\sin\alpha_k)$. Periodicity of the solution imposes a condition between this angle and the ratio of the domain lengths, namely $\tan\alpha_k = L_y/L_x$. Notice that the perpendicular component of the magnetic field is related to the $B_x,B_y$ components by $B_\perp = B_y\cos\alpha_k - B_x\sin\alpha_k$. Such setup admits an analytical, stationary solution, consisting in the advection of the magnetic field along the domain diagonal, with a crossing time $t_{\rm CT}=L_x/\cos\alpha_k$. We set a domain size $L_x=L_y=2$, corresponding to $\alpha_k = \pi/4$ and $t_{\rm CT}=2\sqrt{2}$, and use the ideal EoS with $\Gamma=5/3$.

First we explore the different reconstruction methods introduced in Section~\ref{nonsmooth_FD} (i.e., PPM, FDOC, MP5 and several flavors of WENO), by evolving the Alfven wave in a single mesh with five different resolutions, corresponding to $N=16,32,64,128,256$ points in each direction. All these cases use the conservative discretization, Eq.~(\ref{conservative_discretization}), in combination with the LLF flux formula, Eq.~(\ref{LLF}). The time integration is performed by using the 4$^{\rm th}$-order RK schema with a time-step $\Delta t = \Delta x/(4 \sqrt{2})$, low enough to satisfy the CFL condition and to ensure that the discretization errors are dominated by the spatial terms.
The solutions are evolved
up to 3 crossing times and then we verify the relative error and the convergence rate of the different methods. For each simulation, the relative error is calculated by integrating over the entire domain the L1-norm of the relative difference between the numerical solution of $B_x$ and the analytical one. We checked that the errors calculated over other magnetic field components behave in the same way and that, for each method and resolution, the relative error accumulates and grows linearly with the number of cycles.
\begin{figure}[h]
	\includegraphics[width=8cm]{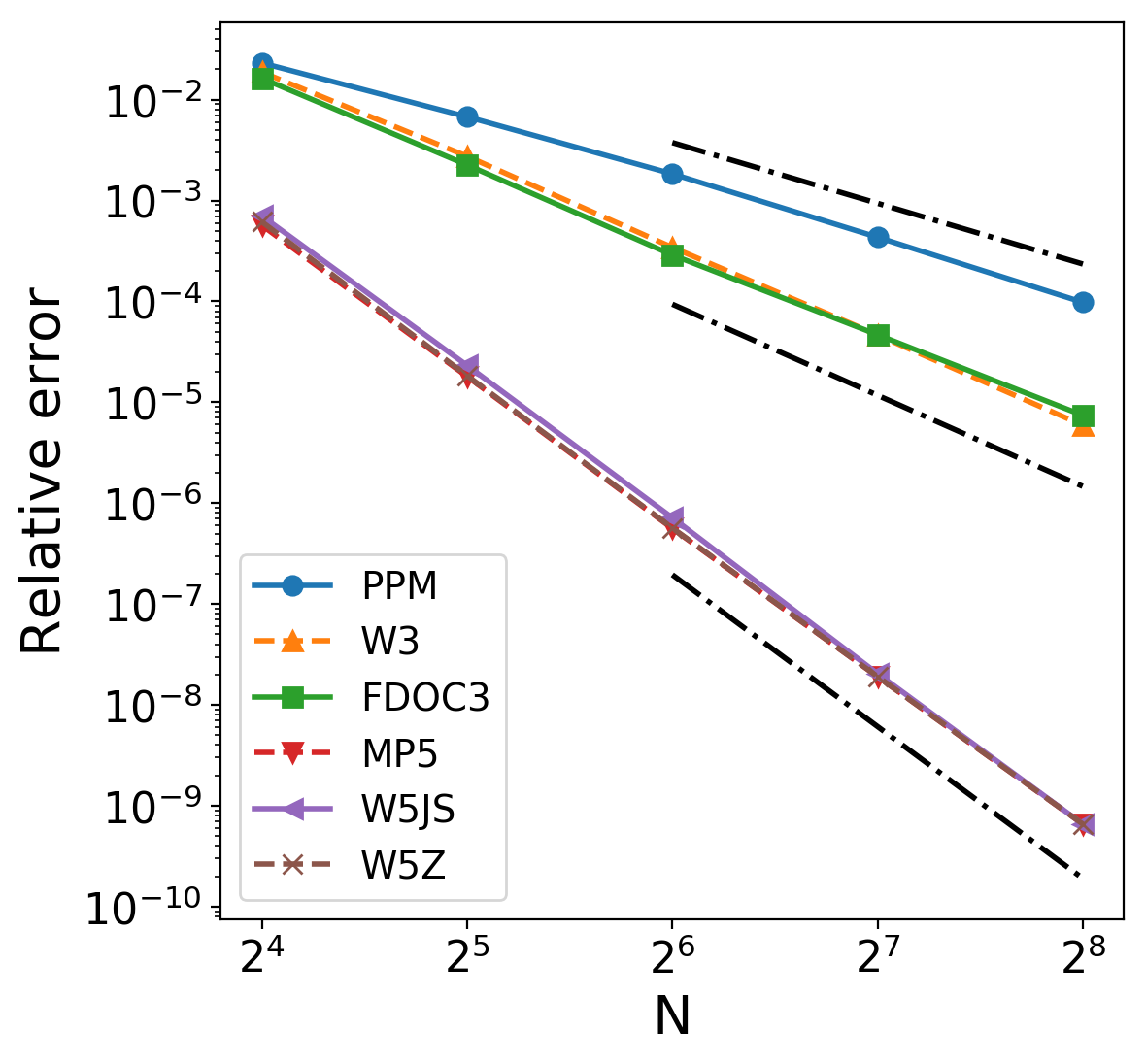}
    \includegraphics[width=8cm]{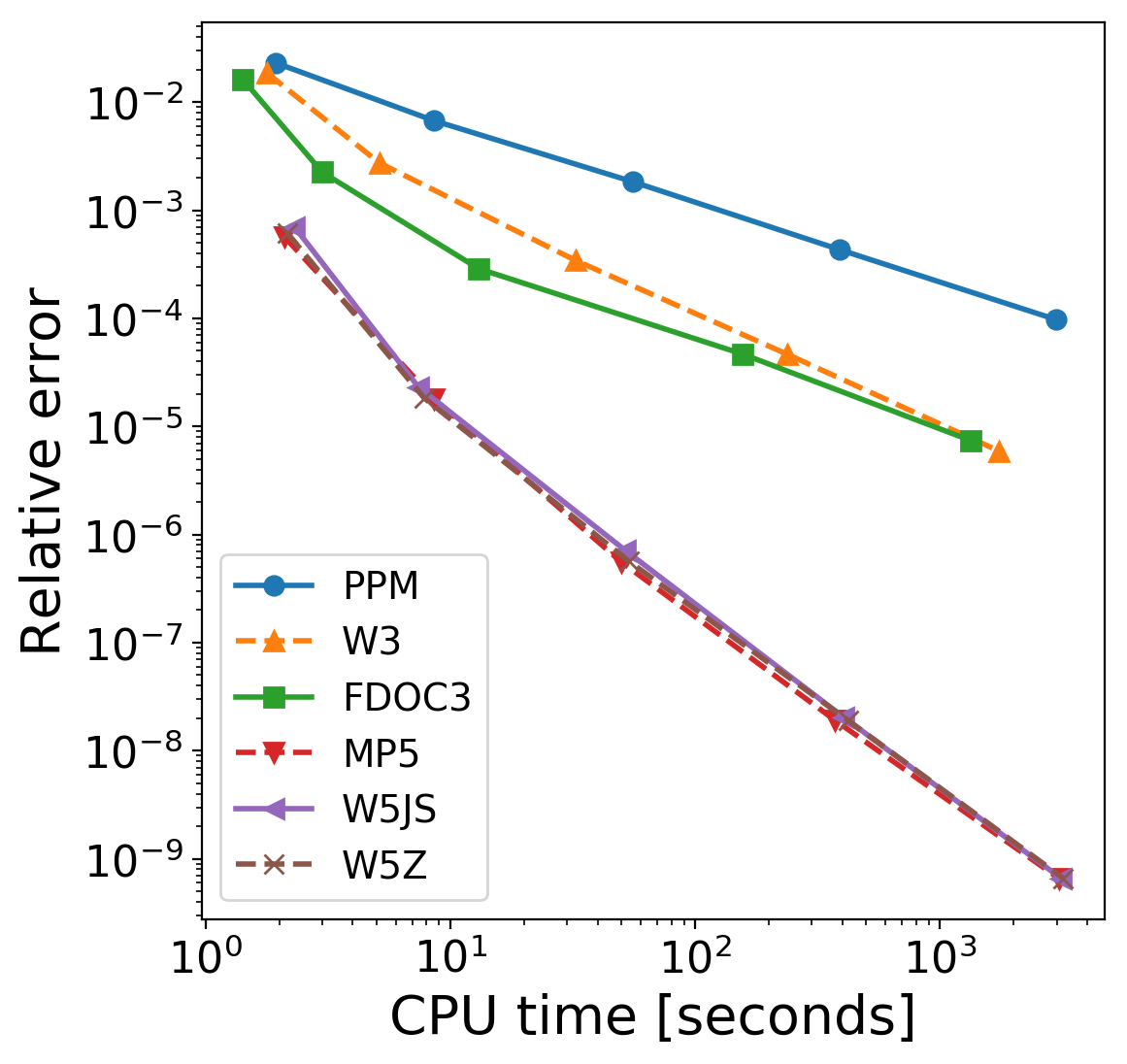}
	\caption{{\em CP Alfven waves in unigrid}. Relative error for different methods as a function of the number of points (top) and CPU time (bottom). The relative error is evaluated between the numerical and the analytical solution of $B_x$ after 3 crossing times (i.e., $t=6\sqrt{2}$), as in Eq.~\ref{def:relative_error}. Clearly, the $5^{\rm th}$-order schemes achieve a smaller error for a given number of points with less CPU time (WENO5Z and MP5 lines almost overlap). The slopes of the black dashed-dotted lines represent, from above to below, the nominal $2^{\rm nd}$, $3^{\rm rd}$ and $5^{\rm th}$ convergence orders. }
	\label{figure:alfven_errors}
\end{figure}
The relative errors of the solution after exactly 3 crossing times are displayed in Fig.~\ref{figure:alfven_errors}. One can clearly see that all the reconstruction methods tested here behave as predicted by their nominal convergence rate: WENO5-JS, WENO5-Z and MP5 show the same $5^{\rm th}$-order convergence, WENO3 and FDOC3 
converge at $3^{\rm rd}$-order and PPM converges only at $2^{\rm nd}$-order 
The relative errors for different resolutions allow us to note that, among the $5^{\rm th}$-order methods, WENO5-Z shows a slight improvement in accuracy, compared to WENO5-JS and MP5.
Besides the accuracy
it is also important to measure the computational cost. The CPU time~\footnote{These tests are performed in a single processor, on a desktop DELL XPS computer, Processor Intel Core i7-7700 CPU, 3.60 GHz.} required to achieve a given accuracy for the different methods is displayed at the bottom panel of Fig.~\ref{figure:alfven_errors}. Note that with the lowest resolution ($N=16$), the CPU time is dominated by the initial data setup, thus it is less dependent on method. We have verified that, for our {\it Simflowny}-generated code and setup, the CPU times here shown stochastically vary by about $5\%-20\%$ depending on the case. Our results indicate that $5^{\rm th}$-order methods are much more efficient than $3^{\rm rd}$-order ones, with a slight preference for WENO5-Z and MP5 over WENO5-JS.

\begin{figure}[h]
	\includegraphics[width=8cm]{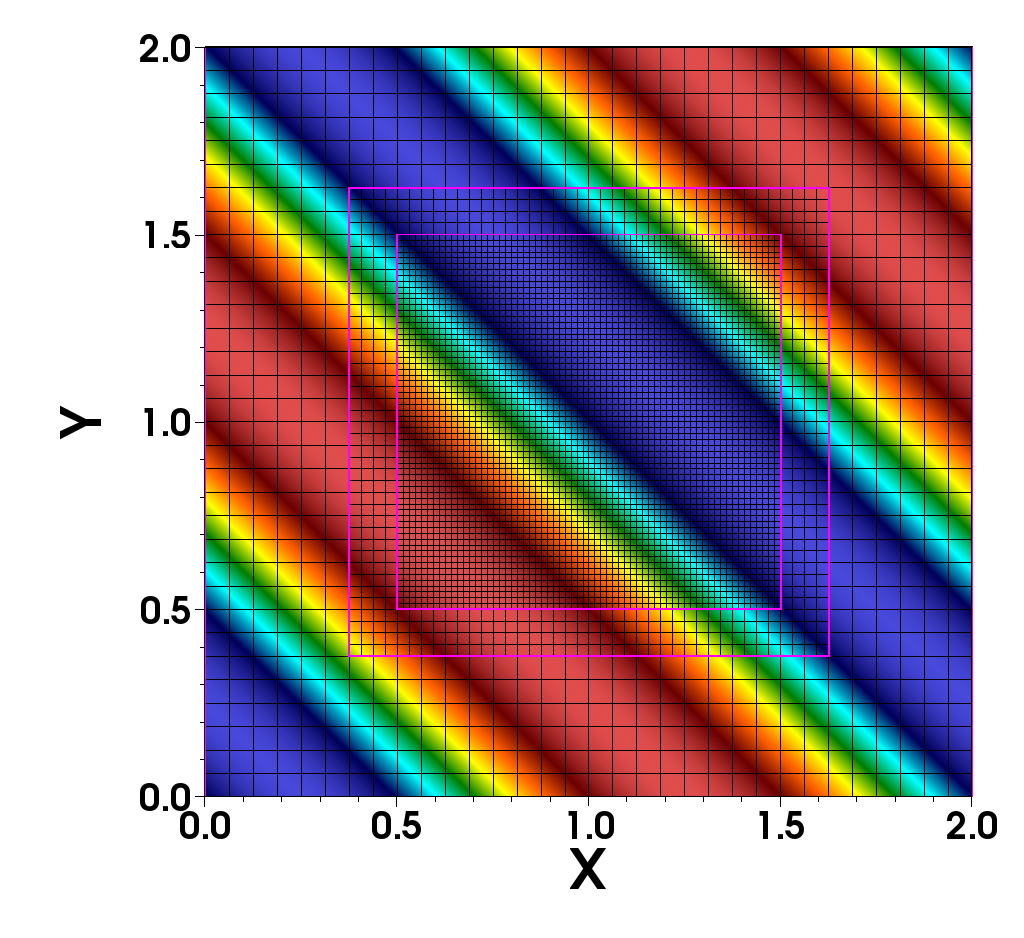}
	\caption{{\em CP Alfven waves with FMR.} Snapshot of $B_z$ component at $t=0.53$ and mesh (drawn in black), for the case with a coarse mesh of 32x32 points, and two levels of refinement with a Ratio 2 each.}
	\label{figure:alfven_visualized}
\end{figure}

\begin{figure}[h]
	\includegraphics[width=8cm]{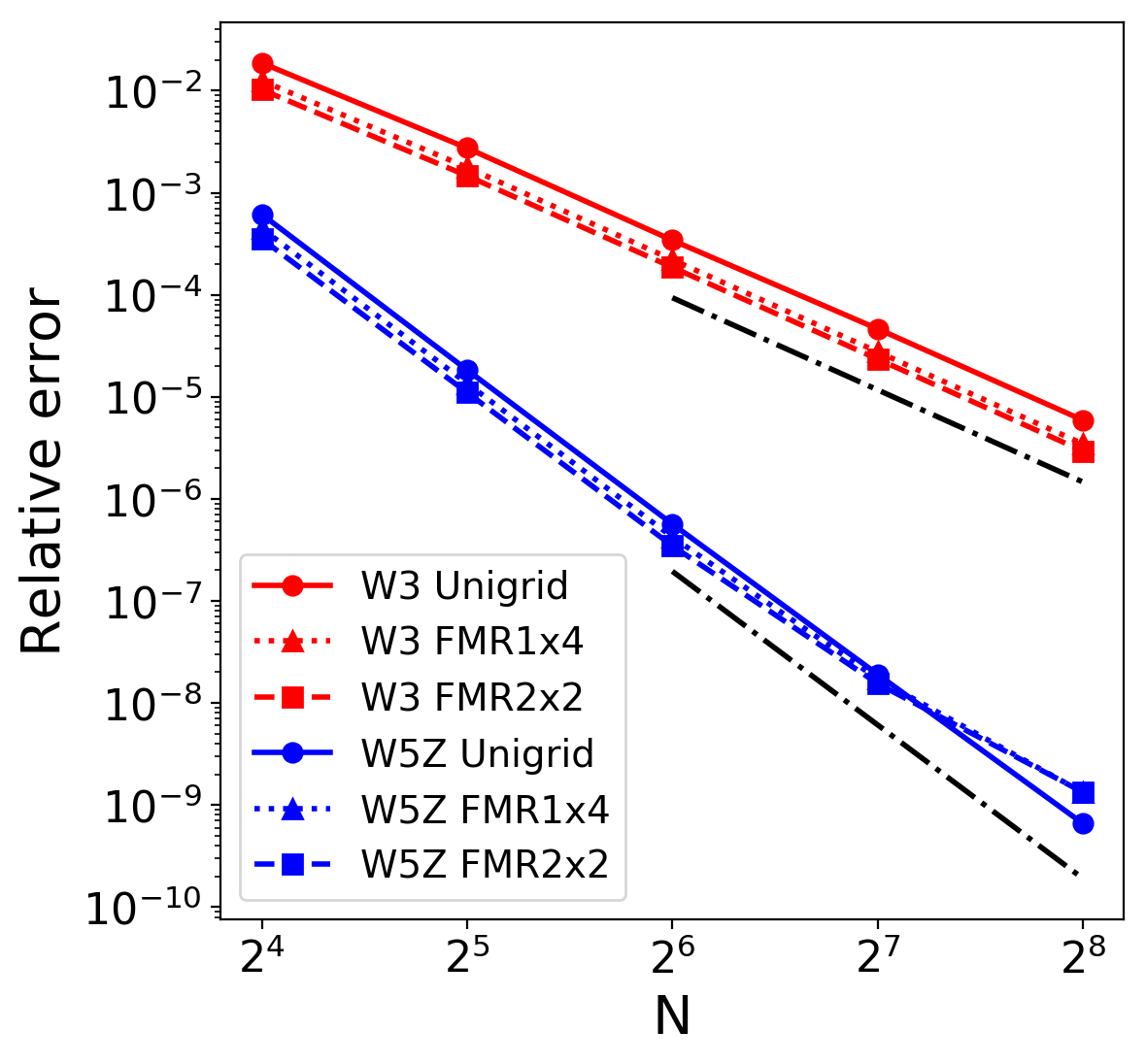}
	\caption{{\em CP Alfven waves with FMR.} Comparison of relative errors of $B_x$ after 3 cycles without/with FMR for WENO3 and WENO5Z, as a function of the number of points $N$ in each direction. The slopes of the black dash-dotted lines represent, from above to below, the nominal $3^{\rm rd}$ and $5^{\rm th}$ convergence orders.} 
	\label{figure:alfven_errors_fmr}
\end{figure}

Next we study the performance of these schemes on refined meshes by evolving the Alfven wave with FMR. We have considered again two different grid setups with additional levels covering the centered half of the domain. The first setup includes two refined grid levels, in addition to the coarse one, with a refinement ratio of two between each level (Ratio 2/2). The $B_z$-component of the solution is displayed in Fig.~\ref{figure:alfven_visualized} within this grid configuration. The second setup has only one refined grid level but with a mesh located at $[0.5:1.5]\times[0.5:1.5]$, with a refinement ratio of four (Ratio 4).

Note that, in a realistic case, this choice of FMR is not computationally convenient, because the solution is propagating in and out from the refined region. This implies that the error, calculated on the main mesh, is dominated by the non-refined region. However, this test is useful to prove that the convergence order is maintained for all the tested methods and no numerical artifacts appear, as it is shown in Fig.~\ref{figure:alfven_errors_fmr} for WENO3 and WENO5-Z. This confirms that the time refinement strategy already tested for the scalar wave equation can be extended successfully to more challenging models.

\subsubsection{Magnetic shock tube}

We further test our code capabilities to capture shocks through a non-smooth MHD problem: the Brio and Wu 1D magnetic shock tube \cite{Brio:1988}. The initial data is given by a constant value of $B_x=0.75$,  $B_z=v_i=0$, and a sharp jump on the other fields, defined by a left and right state:
\begin{eqnarray}
  \rho=1 ~,~ p=1~,~ B_y=1  ~~ && {\rm if}~~ x \leq 0.5 \\
  \rho=0.125~,~ p=0.1~,~ B_y=-1  ~~&&  {\rm if}~~ x>0.5  \nonumber
\end{eqnarray}
We employ the ideal gas EoS with $\Gamma=2$. This problem is the magnetic extension of the classical hydrodynamical Sod shock tube and is particularly challenging for values of magnetic pressure comparable to the fluid pressure, like in the setup here proposed. 

\begin{figure}[h]
	\includegraphics[width=9cm]{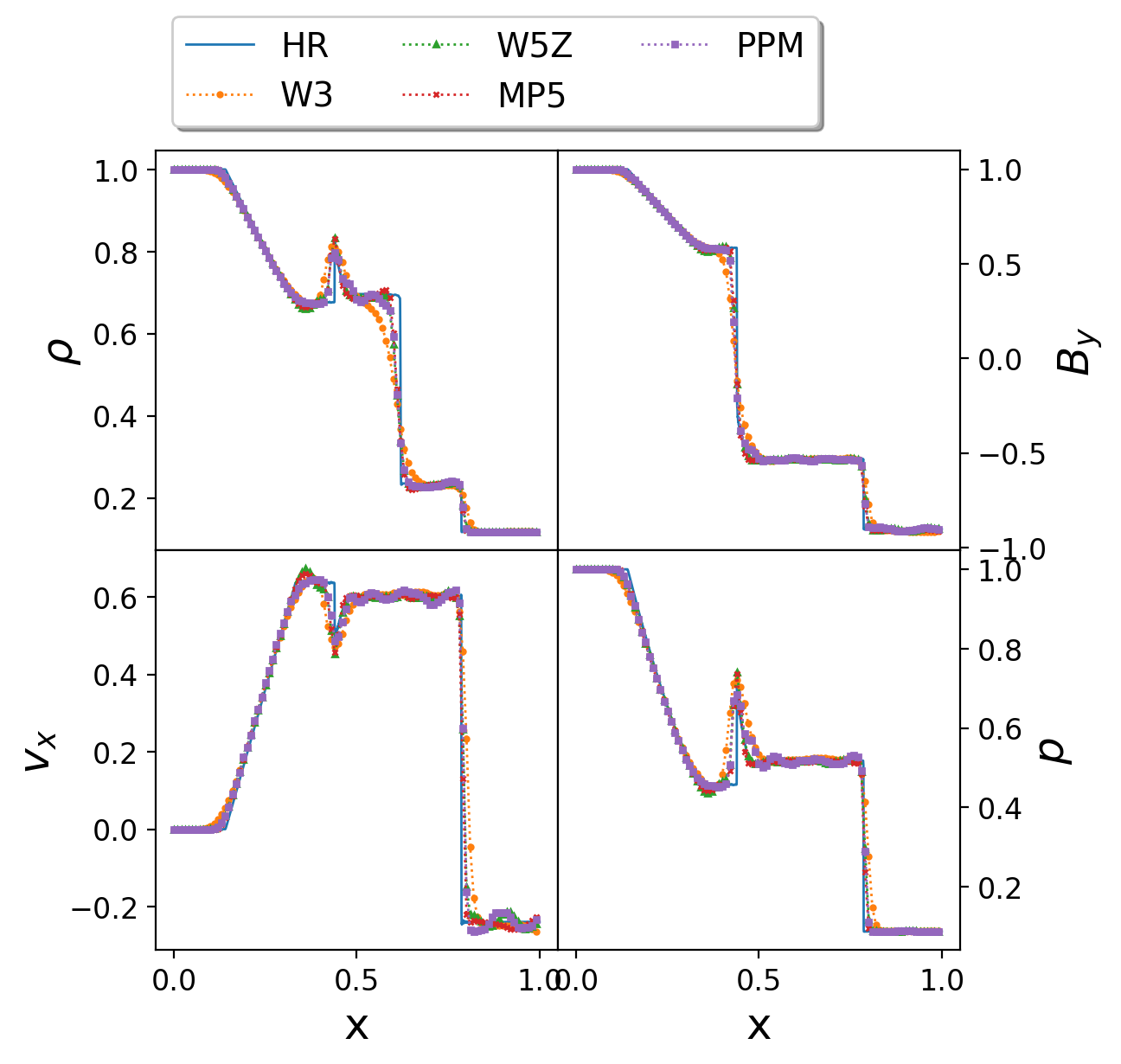}
	\caption{{\em Brio \& Wu shock tube test in unigrid.} Profiles of $\rho$, $B_y$, $v_x$ and $p$, with different reconstruction methods at $t=0.2$ by using $N=100$. The exact solution (solid line, ``HR''), has been evaluated by running the same problem with PPM and $N=4000$.} 
	\label{figure:magshock_profiles}
\end{figure}

We focus for this comparison on some popular reconstruction schemes in the presence of shocks, specifically PPM, WENO3, WENO5-Z and MP5. Our simulations are performed in a 1D channel with $N=50,100,200,400$ points along the x-direction and using a time step $\Delta t=0.2\, \Delta x$. We compare the results with the exact solution (solid line, ``HR'' in figures), evaluated by running the same problem with PPM reconstruction and $N=4000$. We find results quantitatively consistent with other tested codes~\citep{DelZanna:2007,Stone:2008}\footnote{See also the Athena tests webpage:\\ {\tt https://www.astro.princeton.edu/$\sim$jstone/Athena/tests/}}. Hereafter we analyze the results at $t=0.2$, when the solution has fully develop its complex profile, including the propagation of shock and rarefaction waves. In Fig.~\ref{figure:magshock_profiles} the profiles of $\rho$, $B_y$, $v_x$ and $p$ are displayed, obtained with different methods for $N=100$. Note that PPM, as it is well known, is able to attain a satisfactory accuracy for non-smooth solutions, similar to the highest order method, MP5 and WENO5-Z, and much better than WENO3. The velocity component $v_x$ shows the largest mismatch with respect to the exact solution, with oscillations visible even for $N=400$ within all methods. 

\begin{figure}[h]
	\includegraphics[width=9cm]{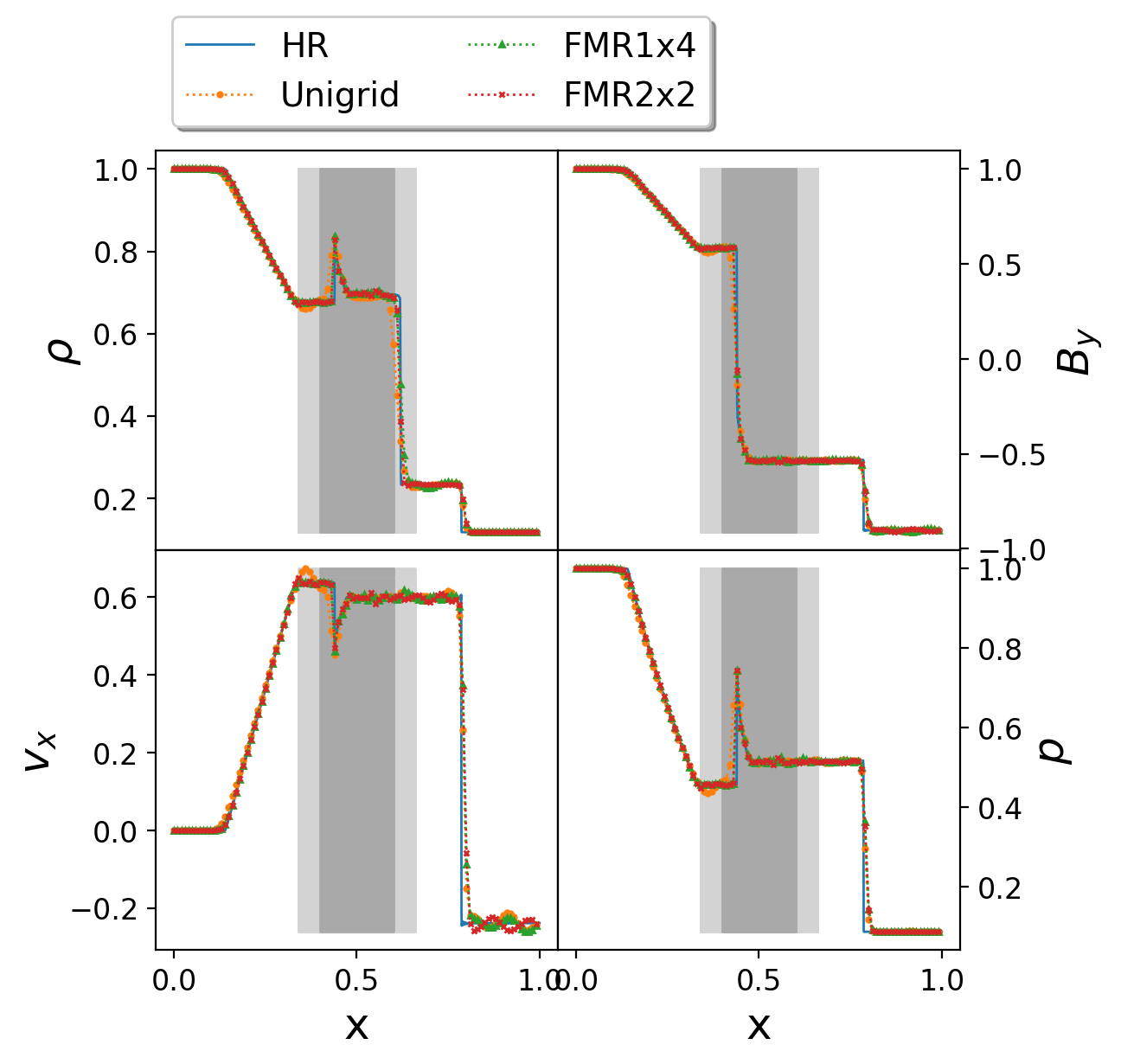}
	\caption{{\em Brio \& Wu shock tube test with FMR.} Same as~\ref{figure:magshock_profiles} only for the reconstruction method WENO5-Z, and comparing unigrid, Ratio 2/2 FMR and Ratio 4 FMR results.}
	\label{figure:magshock_profiles_FMR}
\end{figure}

We repeat the same tests with FMR, considering again two different grid structures: one with two refinement levels and a Ratio 2/2, the first level covering the region $x\in[0.34,0.66]$ and $x\in[0.4,0.6]$ the second. The other has only one refinement level with a ratio 4 covering the region $x\in[0.4,0.6]$. These refined regions are entirely crossed before $t\sim 0.2$ by both the rarefaction wave propagating to the left, and the shock front moving to the right. In Fig.~\ref{figure:magshock_profiles_FMR} we show the same profiles as in the previous figure, comparing the solutions with and without FMR. The refined areas with resolution factors of 2 and 4 compared to the coarse grid are indicated with light and dark grey, respectively. Note that the profiles with low resolution are presented in order to visually appreciate the differences, which are almost indistinguishable by eye for $N\geq 200$.

\begin{figure}[h]
	\includegraphics[width=8cm]{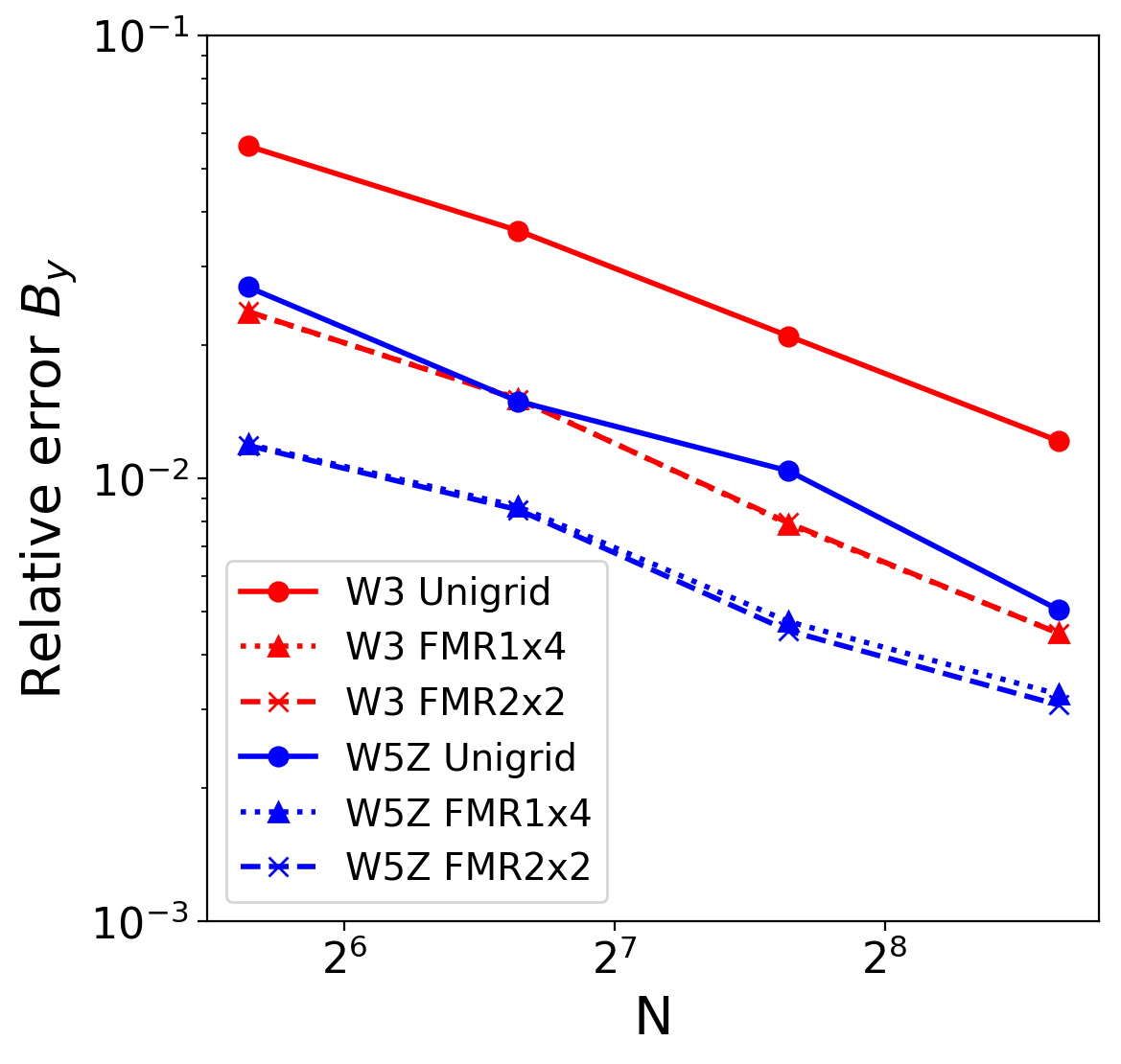}
	\caption{{\em Brio \& Wu shock tube test with FMR.} Relative L1-errors of $B_y$ at $t=0.2$ as a function of resolution by using WENO3 and WENO5-Z with different multi-grid structures. Note that the FMR results with ratio 2/2 and 4 almost overlap.} 
	\label{figure:magshock_errorsN_FMR}
\end{figure}

A quantitative assessment of the gain in accuracy is shown in Fig.~\ref{figure:magshock_errorsN_FMR}, where the comparison of relative errors of $B_y$ for WENO3 and WENO5-Z is displayed for the setups with and without FMR as a function of number of points at $t=0.2$, and taking as a reference solution a high-resolution run with $N=4000$ performed with PPM. Clearly, the addition of refinement grids with different resolution ratios does not degrade the convergence order of the numerical scheme. Furthermore, higher resolution improves the accuracy and it does not introduce any spurious solution.

The conclusions of these MHD tests can be summarized as follow: (i) our implementation of the HRSC methods and monotonic reconstructions are correct, (ii) the BOR sub-cycling strategy for FMR/AMR preserves the accuracy of non-smooth solutions despite the use of an apparently simple Lagrangian interpolation for the prolongation step, and (iii) non-smooth solutions with resolution ratios larger than 2 between consecutive grid levels are equivalent on accuracy to those with ratio 2.


\section{Einstein equations}

This section focuses on the implementation of the conformal and covariant Z4 formulation~\cite{Bona:2003} of the Einstein Equations to study the gravitational wave radiation produced by the merger of black holes and other compact objects. Here only a short summary of the system of equations will be presented, deferring to~\cite{Alic:2012,Bezares:2017,Palenzuela:2017} for full details and numerical applications.
The covariant CCZ4 formalism can be written as
\begin{eqnarray}
   R_{ab} &+& \nabla_a Z_b + \nabla_a Z_b   =
   8\pi \, \left( T_{ab} - \frac{1}{2}g_{ab} \,\trT \right) \nonumber \\
   &+& \kappa_{z} \, \left(  n_a Z_b + n_b Z_a - g_{ab} n^c Z_c \right).
\end{eqnarray}
where $R_{ab}$ is the Ricci tensor associated to the metric $g_{ab}$, $T_{ab}$ is the stress-energy tensor associated to the matter content, $n_a$ is the normal to the spatial hypersurfaces and $Z_a$ is a four-vector introduced to enforce dynamically the energy-momentum constraints in a timesacle $1/\kappa_{z}$. 
Notice that we have chosen geometric units such that $G=c=1$ and we adopt the convention where roman indices a,b,c,... denote spacetime components (i.e., from 0 to 3), while i,j,k,... denote spatial ones.

These covariant equations can be written as an hyperbolic evolution system
by means of the $3+1$ decomposition, which split the spacetime tensors and equations into their space and time components. The line element
can be decomposed as
\begin{equation}
  ds^2 = - \alpha^2 \, dt^2 + \gamma_{ij} \bigl( dx^i + \beta^i dt \bigr) \bigl( dx^j + \beta^j dt \bigr), 
\label{3+1decom}  
\end{equation}
where $\alpha$ is the lapse function, $\beta^{i}$ is the shift vector, and $\gamma_{ij}$ is the induced metric on each spatial foliation, denoted by $\Sigma_{t}$. In this foliation we can define the normal to the hypersurfaces $\Sigma_{t}$ as $n_{a}=(-\alpha,0)$ and the extrinsic curvature $K_{ij} \equiv  -\frac{1}{2}\mathcal{L}_{n}\gamma_{ij}$,  where $\mathcal{L}_{n}$ is the Lie derivative along  $n^{a}$. 

For numerical applications it is more convenient to transform to
a conformal metric $\tilde{\gamma}_{ij}$ with unit determinant. In
terms of a real, positive conformal factor $\chi$, one then obtains  a conformal trace-less extrinsic curvature $\tilde{A}_{ij}$ 
\begin{eqnarray}
\tilde{\gamma}_{ij} &=& \chi\,\gamma_{ij},\\
\tilde{A}_{ij}      &=& \chi\left(K_{ij}-\frac{1}{3}\gamma_{ij} \trK \right),
\end{eqnarray}  
where $\trK=\gamma^{ij}K_{ij}$. For further convenience we can redefine some of the evolved quantities as
\begin{eqnarray}
  \trKhat &\equiv& \trK - 2\, \Theta, \\
  {\hat \Gamma}^i &\equiv& {\tilde \Gamma}^i + 2 Z^{i}/\chi,
\end{eqnarray}  
where $\Theta \equiv - n_{a} Z^{a}$. The final set of evolution
fields is $\{ \chi, {\tilde \gamma}_{ij}, \trKhat, {\tilde A}_{ij}, {\hat \Gamma}^i, \Theta  \}$. The evolution equations for these fields can be written as~\cite{Bezares:2017}:

\begin{widetext}
\begin{eqnarray}
\partial_t {\tilde \gamma}_{ij} 
    & =& \beta^k \partial_k {\tilde \gamma}_{ij} + {\tilde \gamma}_{ik} \, \partial_j \beta^k 
    + {\tilde \gamma}_{kj} \partial_i \beta^k - {2\over3} \, {\tilde \gamma}_{ij} \partial_k \beta^k
    - 2 \alpha \Bigl( {\tilde A}_{ij}  - \frac{1}{3} {\tilde \gamma}_{ij}\, tr {\tilde A} \Bigr) -  \frac{\kappa_{c}}{3}\,\alpha\tilde{\gamma}_{ij}\ln\tilde{\gamma}, \label{syseq1}
\\
\partial_t {\tilde A}_{ij} 
    & =& \beta^k \partial_k{\tilde A}_{ij} + {\tilde A}_{ik} \partial_j \beta^k 
     + {\tilde A}_{kj} \partial_i \beta^k - {2\over3} \, {\tilde A}_{ij} \partial_k \beta^k - \,\frac{\kappa_{c}}{3}\,\alpha\,\tilde{\gamma}_{ij}
     \,tr \tilde{A}
\\
& +& \chi \, \Bigl[ \, \alpha \, \bigl( {^{(3)\!}R}_{ij} + \nabla_i Z_j + \nabla_j Z_i 
    - 8 \pi \, S_{ij} \bigr)  - \nabla_i \nabla_j \alpha \, \Bigr]^{\rm TF} 
 + \alpha \, \Bigl( tr {\hat K} \, {\tilde A}_{ij} - 2 {\tilde A}_{ik} {\tilde A}^k{}_j \Bigr), 
\nonumber \\
\partial_t \chi & =& \beta^k \partial_k \chi 
+ {2\over 3} \, \chi \, \bigl[ \alpha (tr {\hat K} + 2\, \Theta) - \partial_k \beta^k  \bigr], 
\\
\partial_t tr {\hat K} 
    & =&  \beta^k \partial_k tr {\hat K} 
       - \nabla_i \nabla^i \alpha
        + \alpha \, \left[ {1 \over 3} \bigl( tr {\hat K} + 2 \Theta \bigr)^2 
       + {\tilde A}_{ij} {\tilde A}^{ij} + 4\pi  \bigl(\tau + tr S\bigr)
       + \kappa_z  \Theta \right]  
\nonumber \\
       &+& 2\, Z^i \nabla_i \alpha, 
\\
\partial_t \Theta 
    & =&  \beta^k \partial_k \Theta + {\alpha \over 2} \left[ {^{(3)\!}R} + 2 \nabla_i Z^i
   + {2\over3} \, tr^2{\hat K} + {2\over3} \, \Theta \Bigl( tr {\hat K} - 2 \Theta \Bigr)
   - {\tilde A}_{ij} {\tilde A}^{ij}  \right] - Z^i \nabla_i \alpha 
\nonumber \\
   &-& \alpha \, \Bigl[ 8\pi  \, \tau  + 2\kappa_z  \, \Theta \Bigr], 
\\   
\partial_t {\hat \Gamma}^i 
    & =& \beta^j \partial_j {\hat \Gamma}^i - {\hat \Gamma}^j \partial_j \beta^i 
    + {2\over3} {\hat \Gamma}^i \partial_j \beta^j + {\tilde \gamma}^{jk} \partial_j \partial_k \beta^i
   + {1\over3} \, {\tilde \gamma}^{ij} \partial_j \partial_k \beta^k \nonumber
\\
& -& 2 {\tilde A}^{ij} \partial_j \alpha + 2\alpha \, \Bigl[ {\tilde \Gamma}^i{}_{jk} {\tilde A}^{jk}
  - {3 \over 2 \chi} \, {\tilde A}^{ij} \partial_j \chi 
  - {2\over3} \, {\tilde \gamma}^{ij} \partial_j tr{\hat K} - 8\pi \, {\tilde \gamma}^{ij} \, S_i \Bigr] \nonumber
\\ 
& +& 2 \alpha \, \left[- {\tilde \gamma}^{ij} \left( {1 \over 3}\partial_j \Theta 
  + {\Theta \over \alpha} \, \partial_j \alpha \right) 
  - {1 \over \chi} Z^i \left( \kappa_z + {2\over 3} \, (tr{\hat K} + 2 \Theta) \right) \right],   
   \label{syseq2}
\end{eqnarray}
\end{widetext}
where $\{\tau \equiv n_a \, n_b \, T^{ab}, S_i \equiv -n^a T_{ai}, S_{ij} \equiv  T_{ij} \}$ are the projections of the stress-energy tensor, ${^{(3)\!}R_{ij}}$ is the spatial three-dimensional Ricci tensor associated to the metric ${\gamma}_{ij}$ and the expression $[\ldots]^{\rm TF}$ indicates the trace-free part with respect to the metric $\tilde{\gamma}_{ij}$. Notice that there are two sets of constraints in the system; the physical energy and momentum constraints and the conformal constraints,
which are enforced dynamically during the evolution by setting $\kappa_z, \kappa_c >0$.
Again, for black holes, a successful choice is to set $\kappa_z=\kappa_c=1/M$. 
The non-trivial terms inside this expression can be written as:
\begin{eqnarray}
{^{(3)\!}R}_{ij} & + & 2  \nabla_{(i} Z_{j)} 
   = {^{(3)\!}{\hat R}}_{ij} + {\hat R}^\chi_{ij} ,
\nonumber \\
{\hat R}^{\chi}_{ij} & = & {1 \over 2 \chi} \, \partial_i \partial_j \chi 
    - {1 \over 2 \chi} \, {{\tilde \Gamma}^k}_{ij} \partial_k \chi 
\nonumber \\    
   & -& {1 \over 4 \chi^2} \, \partial_i \chi \partial_j \chi 
    + {2 \over \chi^2} Z^k {\tilde \gamma}_{k(i} \partial_{j)} \chi 
\nonumber\\
 &+& {1 \over 2 \chi }{\tilde \gamma}_{ij} \, \Bigl[ {\tilde \gamma}^{km} \Bigl( {\partial}_k {\partial}_m \chi 
 -  {3\over 2 \chi} \, \partial_k \chi \partial_m \chi \Bigr)
\nonumber \\ 
 & &~~~~~~ - {\hat \Gamma}^k \partial_k \chi \Bigr] ,
\nonumber     \\
{^{(3)\!}{\hat R}}_{ij} &=& - {1\over2} \, {\tilde \gamma}^{mn} \partial_m \partial_n {\tilde \gamma}_{ij}     
        + {\tilde \gamma}_{k(i} \partial_{j)} {\hat \Gamma}^k 
\nonumber
\\        
        &+&  {\hat \Gamma }^k {\tilde \Gamma}_{(ij)k} + {\tilde \gamma}^{mn} 
             \Bigl(  {{\tilde \Gamma}^k}_{mi} {\tilde \Gamma}_{jkn} 
        \Bigr.
\nonumber   \\
         &+&  {{\tilde \Gamma}^k}_{mj} {\tilde \Gamma}_{ikn} +{\tilde \Gamma}^k{}_{mi} {\tilde \Gamma}_{knj}  \Bigr), 
\nonumber \\
 \nabla_i \nabla_j \alpha & = & \partial_i \partial_j \alpha - {{\tilde \Gamma}^k}_{ij} \partial_k \alpha + {1 \over 2 \chi} \Bigl( \partial_i \alpha\, \partial_j \chi \Bigr.
\nonumber \\ 
  &+&  \Bigl. \partial_j \alpha  \, \partial_i \chi
           - {\tilde \gamma}_{ij}\, {\tilde \gamma}^{km}\, \partial_k \alpha\, \partial_m \chi  \Bigr),  
\nonumber
\end{eqnarray}

In order to close the system of equations, coordinate (or gauge) conditions for the evolution of the lapse and shift must be supplied. We use the Bona-Mass\'o family of slicing conditions~\cite{Bona:1995} and the Gamma-driver shift condition~\cite{Alcubierre:2003}, namely 
\begin{eqnarray}
\partial_t \alpha & = &  \beta^i \partial_i \alpha - \alpha^{2} \,f\, tr\hat{K}, \\ 
\partial_t \beta^i & = &  \beta^j \partial_j \beta^i +  \,g \, \, B^{i}, \\
\partial_t B^i & = & \beta^j \partial_j B^i - \eta B^i 
 + \partial_t {\hat \Gamma}^i - \beta^j \partial_j {\hat \Gamma}^{i},
\end{eqnarray}
being $f$ and $g$ arbitrary functions depending on the lapse and the metric, and $\eta$ a constant parameter. For black holes of mass $M$ a common and successful choice is $f=2/\alpha$, $g=3/4$ and $\eta=2/M$.

On the following problems we will consider radiative boundary conditions.
The main part of the radiative boundary conditions assumes that there is an outgoing radial wave with some speed $v$,
\begin{equation}
   U =  U_{\infty} + \frac{f(r-v t)}{r}
\end{equation}
where $U$ is any of the tensor components of the evolved fields, $U_{\infty}$ its value at infinity and $f$ a spherically symmetric perturbation. Notice that $\{U_{\infty},v\}$ depend on the particular field. The time derivative can be written as
\begin{equation}
  \partial_t U = - v^i \partial_i U 
              - v \frac{(U - U_{\infty})}{r}
\end{equation}
where $v^i = v x^i/r$ and $\partial_i$ is evaluated
using centered finite differencing where possible and one-sided elsewhere.

The initial data for multiple spinning and boosted binary black holes can be written as a function of a conformal factor $\psi$, namely~\cite{Cook:2000}
\begin{eqnarray}
  trK &=& 0 ~~,~~ {\tilde \gamma}_{ij} = \eta_{ij}
  \rightarrow \chi = \psi^{-4} ~~,~~ \nonumber \\
  \alpha&=& \psi^{-2} ~~,~~
  \beta^i = B^i = {\tilde \Gamma}_i = 0
\end{eqnarray}
This conformal factor can approximately satisfy the energy constraint when it is defined as a superposition of $M$-single black hole solutions
\begin{eqnarray}
  \psi = 1 + \sum_{A=1}^{M} \frac{m_A}{2 r_A} 
  \label{conformal_factor}
\end{eqnarray}
where $r_A$ is the distance to the location of the black hole $(A)$ with bare mass $m_A$. 
Although better numerical approximations can be found by solving the elliptic energy constraint, for our simple test we will just use the above-mentioned superposition.

The momentum constrain is identically satisfied
with the Bowen-York solution 
\begin{eqnarray}
  {\tilde A}_{ij} &=& \sum_{A=1}^{M}
   \frac{3}{2 r_A^2 \psi^6} 
 \left[ P^{(A)}_i n^{(A)}_j + P^{(A)}_j n^{(A)}_i 
 \nonumber \right. \\
      &-& \left. \left(\eta_{ij} - n^{(A)}_i n^{(A)}_j \right) P_{(A)}^k n^{(A)}_k \right] 
\nonumber \\
       &+& \sum_{A=1}^{M}
   \frac{3}{r_A^3 \psi^6} 
        \left[ \epsilon_{kil} S_{(A)}^l n_{(A)}^k n^{(A)}_j +
               \epsilon_{kjl} S_{(A)}^l n_{(A)}^k n^{(A)}_i  \right]
\nonumber               
\end{eqnarray}
where ${n_i}^{(A)} = x_i^{(A)}/r_A$ is a radial vector not to be confused with the normal to the hypersurfaces. 
Notice that $P^{(A)}$ denotes the momentum $m_A V_A$ of the black hole and $S^{(A)}$ its spin components.

This model will be solved by using the $4^{\rm th}$-order RK with $4^{\rm th}$-order centered spatial derivatives and $6^{\rm th}$-order Kreiss-Oliger dissipation. We will use FMR/AMR with a quintic polynomial Lagrange spatial interpolator and BOR strategy for the sub-ciclying in time. Therefore, the resulting scheme should be $4^{\rm th}$-order accurate both in time and space.

\subsection{Single BH}

Our first test with the Einstein equations is   a single
spinning black hole solution, that allow us to test
a non-trivial non-linear solution together with the
scheme to compute the gravitational waves. Our initial
data, based on the Bowen-York extrinsic curvature for a
single black hole, can be considered as a Kerr black hole plus 
some gravitational radiation that will either propagate
to infinity or accrete into the black hole.
We choose a mass $M=1$ and spin $a=J/M^2=0.2$, following~\cite{Brugmann:2008}.
We will evolve this spinning BH on a grid with several levels
of refinement and two different grid structures as described in Table~\ref{table:mesh248}.

\begin{table}
	\begin{tabular}{| l | l | l | | l | l |}
	     \hline
    &\multicolumn{2}{|c||}{Ratio 2}& \multicolumn{2}{|c|}{Variable Ratio} \\ \hline	     
		Level & $\Delta x_{0}/\Delta x$ & Domain & $\Delta x_{0}/\Delta x$ & Domain \\ \hline
		0 & 1 & [$-120$, $120$] & 1 & [$-120$, $120$] \\ \hline
		1 & 2 & [$-60$, $60$] & 2 & [$-60$, $60$]\\ \hline
		2 & 4 & [$-30$, $30$] & 4 & [$-30$, $30$] \\ \hline
		3 & 8 & [$-15$, $15$] & 16 & [$-15$, $15$] \\ \hline
		4 & 16 & [$-7.5$, $7.5$] & 128 & [$-1.875$, $1.875$] \\ \hline
		5 & 32 & [$-3.75$, $3.75$]& & \\ \hline
		6 & 64 & [$-1.875$, $1.875$] & &\\ \hline
		7 & 128 & [$-0.9375$, $0.9375$] & &\\ \hline	     
	\end{tabular}
	\caption{{\em Single Black Hole}. Mesh structure for the
	case with ratio 2 in the grid spacing and for the case 
	with variable ratios ranging from 2--8.	}
				\label{table:mesh248}
\end{table}

\begin{figure}[h]
	\includegraphics[width=8cm]{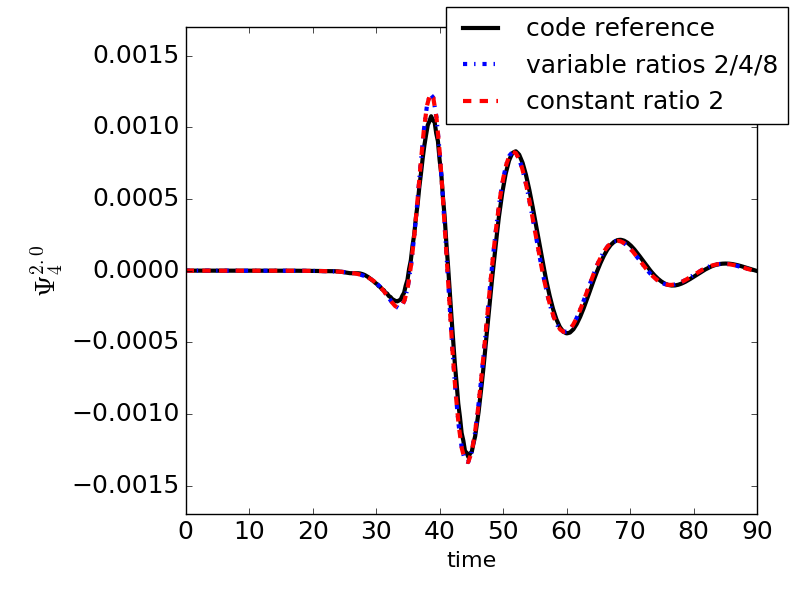}
	\caption{{\em Single black hole}. Gravitational wave at $r=15 $ with two different grid structures, one with constant resolution ratio of 2 and the other with ratios ranging from 2 to 8.
	}
	\label{figure:wave_comparison}
\end{figure}

The emitted GWs are extracted by computing the Newman-Penrose complex scalar $\Psi_4$ in a surface far away from the source, and it is decomposed in a base
of spin-weighted spherical harmonics with $s=-2$ (i.e. see for instance ~\cite{Bishop:2016}). The main gravitational wave mode, corresponding to $l=2,m=0$, is displayed in Fig. \ref{figure:wave_comparison} for the two different grid structures. We can also compare directly with HAD~\cite{Liebling:2002}\footnote{see also the HAD webpage {\tt http://www.had.liu.edu}}, a mature and well-tested code that will serve us as a reference. Although there are some small differences due to the RK integrator (i.e., $4^{\rm th}$-order versus $3^{\rm rd}$-order), the overall agreement is very good. Finally, we can also compare with Fig. 5 from~\cite{Brugmann:2008}, showing
again a very good agreement.

\begin{table}
	\begin{tabular}{| l | l | l | | l | l |}
	     \hline
    &\multicolumn{2}{|c||}{Ratio 2}& \multicolumn{2}{|c|}{Ratio 4} \\ \hline	     
		Level & $\Delta x_{0}/\Delta x$ & Domain & $\Delta x_{0}/\Delta x$ & Domain \\ \hline
		0 & 1 & [$-10$, $10$] & 1 & [$-10$, $10$] \\ \hline
		1 & 2 & [$-5.5$, $5.5$] & 4 & [$-4.85$, $4.85$]\\ \hline
		2 & 4 & [$-4.65$, $4.65$] & 16 & [$-2.5$, $2.5$] \\ \hline
		3 & 8 & [$-2.6$, $2.6$] & 64 & [$-1.25$, $1.25$] \\ \hline
		4 & 16 & [$-2.43$, $2.43$] &  &  \\ \hline
		5 & 32 & [$-1.285$, $1.285$]& & \\ \hline
		6 & 64 & [$-1.233$, $1.233$] & &\\ \hline
	\end{tabular}
	\caption{{\em Single Black Hole}. Mesh structure for the scaling benchmark
	with ratio 2 and Ratio 4 in the grid spacing.	}
	\label{table:mesh_scalability}
\end{table}

\begin{figure}[h]
	\includegraphics[width=7cm]{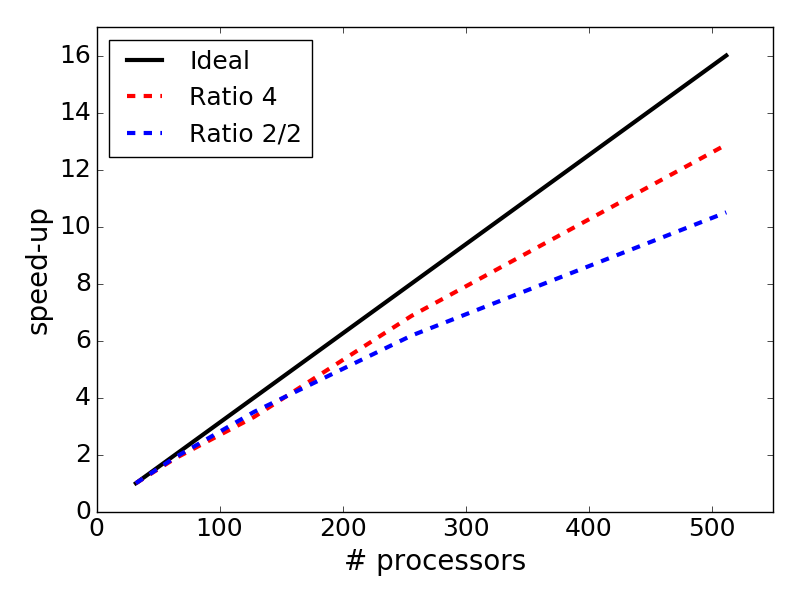} 	\vspace{0.1cm}
	\includegraphics[width=7cm]{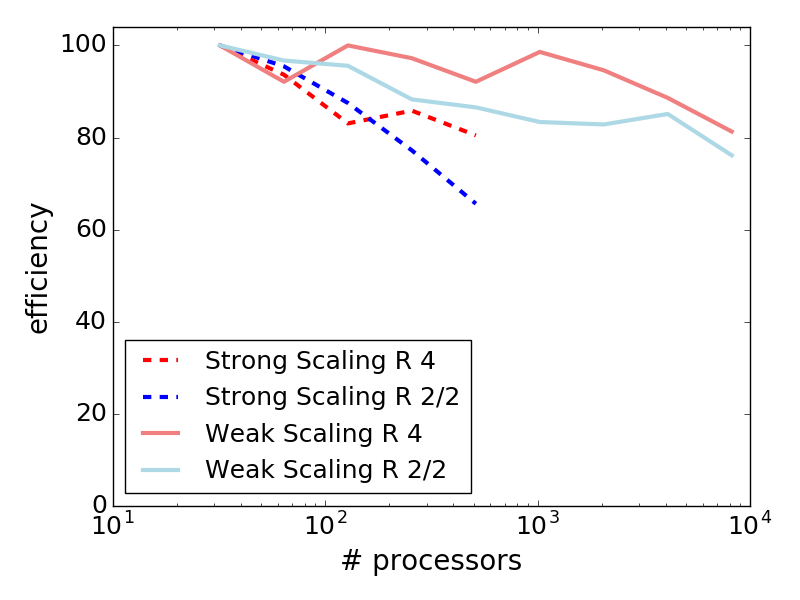}\\
	\caption{{\em Single Black Hole.} Speedup and efficiency for the strong/weak test performed with the CCZ4 formulation of the Einstein equations with several levels of refinement, as described in Table~\ref{table:mesh_scalability}. The time integrator is RK4, we are using $4^{\rm th}$-order space discretization with one-side advective terms and $6^{\rm th}$-order KO dissipation. BOR is being used here as an AMR strategy for the sub-cycling in time.}
	\label{figure:solutionAMR}
\end{figure}

This initial data can also serve us to perform a benchmark on the scalability of the code by using the two different configurations given by Table~\ref{table:mesh_scalability}. The results are displayed in Fig.~\ref{figure:solutionAMR}. The top panel shows that the implementation scales strongly up to a factor of 16 with respect to the minimum number of processors required for this problem. The speed-up (or the efficiency) increases when the resolution ratio is higher than 2, since the load of the fine grids is larger in those situations. The bottom panel shows that the implementation scales quite well (i.e., above $80\%$) at least up to ${\cal O}(10^4)$ processors. Scalability can be increased by using  numerical schemes with small stencils. For instance, SAMRAI has been proven to reach exascale ${\cal O}(10^6)$ for some particular problems. However, low stencil schemes implies low accuracy methods in finite difference and finite volume schemes. Since we are interested not only on scalability but also on efficiency, we will focus our investigations on the other options. 

\subsection{Binary BHs}

Our next test with the Einstein equations involves a binary black hole collision to check the scalability of code. For simplicity we choose a pair of non-spinning identical black holes with a bare mass $m_1=m_2=1/2$, located at $x=\pm 5$ and initially at rest momentum. We will evolve this puncture BH on a grid with several levels of fixed mesh refinement, leaving the last refined grids with adaptive mesh refinement.
We perform this simulation by using again two different
grid structures for the AMR grids: with a constant resolution ratio 2, and with a resolution ratio 4.
The evolution parameters will be identical to the single black hole.  

\begin{figure*}[ht]
	\includegraphics[width=16cm]{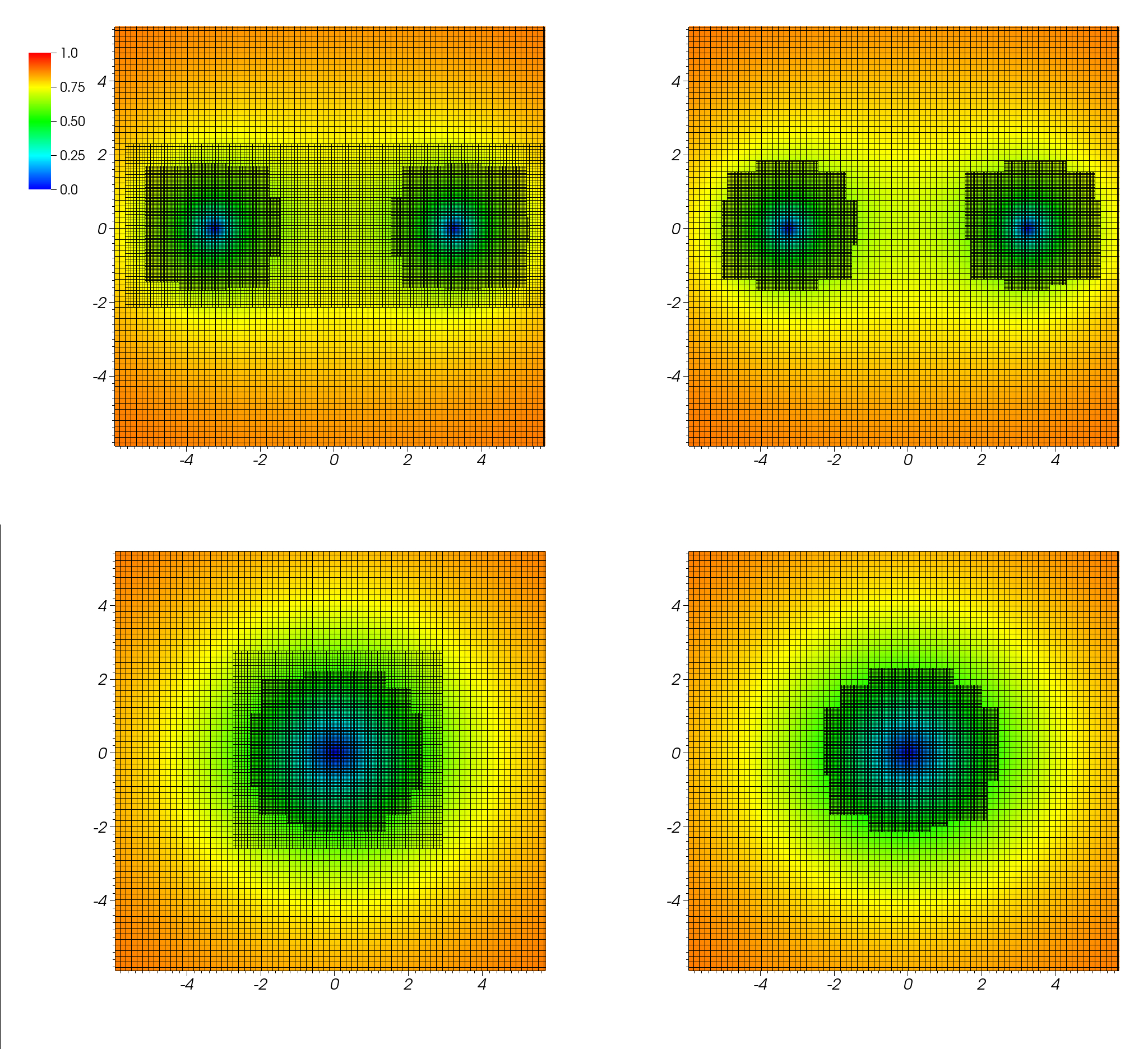}
	\caption{{\em Binary black holes}. Time snapshots of the lapse on the equatorial plane $z=0$, together with the two AMRs, either with constant resolution ratio 2 or with multi-ratios ranging from 2 to 4.
	}
	\label{figure:snapshot_binarybh_2d}
\end{figure*}

\begin{figure}[h]
	\includegraphics[width=7cm]{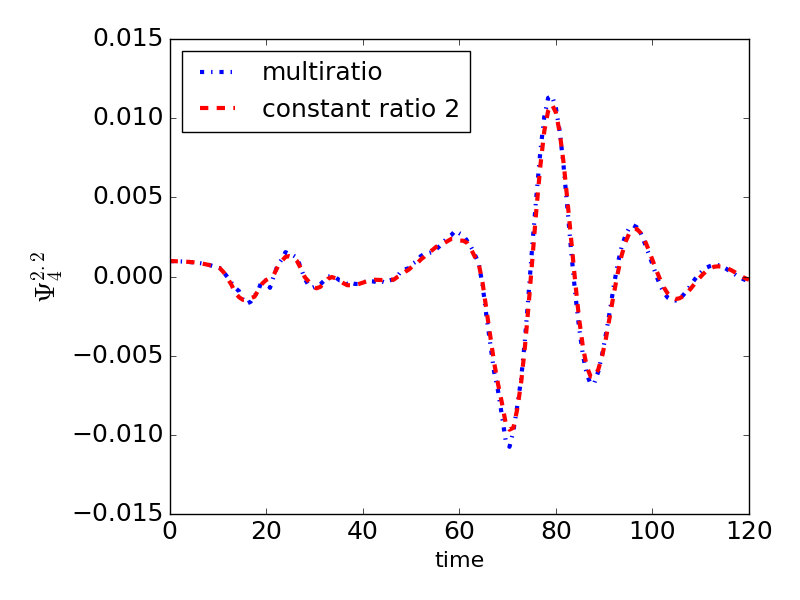}
	\caption{{\em Binary black holes}. Gravitational wave at $r=15 $ with two different grid structures, one with constant resolution ratio of 2 and the other with variable ratios ranging 2 and 4.
	}
	\label{figure:wave_comparison_headon}
\end{figure}

Each black hole feels the gravity produced by 
the other one and move towards each other, leading
to a head-on collision and producing a larger black hole. Some snapshots of the lapse at different times are displayed in Figure~\ref{figure:snapshot_binarybh_2d}. The accelerated motion of the black holes produces
gravitational waves, which can be measured in a 
close spherical surface. We compare the waveforms
obtained with the two different grid structures
to asses that the constant ratio 2 and the
variable ratio lead to comparable solutions. As it can be seen in Figure~\ref{figure:wave_comparison_headon} they are leading to the same results also when AMR is allowed.

\section{Conclusions}

Numerical relativity at exascale level has been out of reach due to the lack of scalability of many traditional approaches, which worked well in the order of a few hundreds-thousands of processors. However, the challenges in multi-messenger astronomy is making scalability a pressing issue. In this paper we have presented the numerical implementation of several computational models by using {\it Simflowny}, a platform to automatically generate efficient parallel code for simulation frameworks. Here we focused on the SAMRAI infrastructure, which has been shown to reach exascale for some specific problems.
We have described in detail the advanced numerical techniques (i.e., spatial discretization schemes, time integrator and AMR strategies for the time refinement) that we implemented in {\it Simflowny} in order to deal with hyperbolic-parabolic systems of equations. 
We believe that the suitable combination of these features (high-order schemes, arbitrary resolution ratios and optimal time sub-cycling strategies) allow us to construct a state-of the art finite-difference code which is fast, efficient, accurate and highly scalable. 

Then, we have considered two different test models to check these numerical techniques and validate the code generation. The first one is the wave equation, that allowed us to test the spatial discretization for smooth solutions and the different sub-cycling in time refinement strategies. The second model is the Newtonian MHD equations, that allowed us to test the spatial discretization for non-smooth solutions and the AMR algorithms in the presence of discontinuities and shocks. Finally, we have implemented the CCZ4 formulation of the Einstein equations to study black hole scenarios. We show that the solutions of our code implementation reproduces the same results as other well-tested codes. We have also studied the performance of our code with weak and strong scaling tests, showing a good scalability at least up to 10,000 processors, which is our current limit of available computational resources. Notice that, since weak scalability is excellent, the strong scalability depends not on the total number of processors but instead on the memory load per processor. This means that the ratio with respect to the minimum number of processors, achieved when the memory load fill the RAM memory of each processor, is the only important factor. Therefore, our code scales strongly above $80\%$ efficiency, for the configuration with Ratio 4, at least up to a factor 16 between the minimum and the maximum number of processors.

We can conclude that our automatically-generated code is production-ready for demanding numerical relativity applications to attain simulations with a larger number of processors than competing codes. Higher-order schemes, together
with advanced time sub-cycling strategies and multi-ratio AMR resolutions, allow our code to not only outperform state-of-the-art finite-difference approaches on scalability but also on accuracy.
More complicated systems, including relativistic magnetized fluids with realistic EoS, require additional specific ongoing developments
that will be reported in future works. This will allow us to simulate binary neutron star mergers in tens of thousands of processors to reach the accuracy required to capture the relevant physical processes associated to the electromagnetic emission.

\subsection*{Acknowledgments} 

We would like to thank Steve Liebling for his useful comments on this manuscript. We also acknowledge support from the Spanish Ministry of Economy, Industry and Competitiveness grants AYA2016-80289-P and  AYA2017-82089-ERC (AEI/FEDER, UE). CP also acknowledges support from the Spanish Ministry of Education and Science through a Ramon y Cajal grant. MB would like to thank CONICYT Becas Chile (Concurso Becas de Doctorado en el Extranjero) for financial support.
We thankfully acknowledge the computer resources at MareNostrum and the technical support provided by Barcelona Supercomputing Center (RES-AECT-2018-1-0003).

%
%
\bibliography{./biblio}

\begin{thebibliography}{53}
\expandafter\ifx\csname natexlab\endcsname\relax\def\natexlab#1{#1}\fi
\expandafter\ifx\csname bibnamefont\endcsname\relax
  \def\bibnamefont#1{#1}\fi
\expandafter\ifx\csname bibfnamefont\endcsname\relax
  \def\bibfnamefont#1{#1}\fi
\expandafter\ifx\csname citenamefont\endcsname\relax
  \def\citenamefont#1{#1}\fi
\expandafter\ifx\csname url\endcsname\relax
  \def\url#1{\texttt{#1}}\fi
\expandafter\ifx\csname urlprefix\endcsname\relax\def\urlprefix{URL }\fi
\providecommand{\bibinfo}[2]{#2}
\providecommand{\eprint}[2][]{\url{#2}}

\bibitem[{\citenamefont{Abbott et~al.}(2016{\natexlab{a}})}]{LIGO:2016blz}
\bibinfo{author}{\bibfnamefont{B.~â.} \bibnamefont{Abbott}}
  \bibnamefont{et~al.} (\bibinfo{collaboration}{Virgo, LIGO Scientific}),
  \bibinfo{journal}{Phys. Rev. Lett.} \textbf{\bibinfo{volume}{116}},
  \bibinfo{pages}{061102} (\bibinfo{year}{2016}{\natexlab{a}}),
  \eprint{1602.03837}.

\bibitem[{\citenamefont{Abbott et~al.}(2016{\natexlab{b}})}]{LIGO:2016nmj}
\bibinfo{author}{\bibfnamefont{B.}~\bibnamefont{Abbott}} \bibnamefont{et~al.}
  (\bibinfo{collaboration}{Virgo, LIGO Scientific}), \bibinfo{journal}{Phys.
  Rev. Lett.} \textbf{\bibinfo{volume}{116}}, \bibinfo{pages}{241103}
  (\bibinfo{year}{2016}{\natexlab{b}}), \eprint{1606.04855}.

\bibitem[{\citenamefont{Abbott et~al.}(2017)}]{LIGO:2017vtc}
\bibinfo{author}{\bibfnamefont{B.~P.} \bibnamefont{Abbott}}
  \bibnamefont{et~al.} (\bibinfo{collaboration}{Virgo}),
  \bibinfo{journal}{Phys. Rev. Lett.} \textbf{\bibinfo{volume}{118}},
  \bibinfo{pages}{221101} (\bibinfo{year}{2017}), \eprint{1706.01812}.

\bibitem[{\citenamefont{{Abbott}
  et~al.}(2017{\natexlab{a}})\citenamefont{{Abbott}, {Abbott}, {Abbott},
  {Acernese}, {Ackley}, {Adams}, {Adams}, {Addesso}, {Adhikari}, {Adya}
  et~al.}}]{LIGO:2017a}
\bibinfo{author}{\bibfnamefont{B.~P.} \bibnamefont{{Abbott}}},
  \bibinfo{author}{\bibfnamefont{R.}~\bibnamefont{{Abbott}}},
  \bibinfo{author}{\bibfnamefont{T.~D.} \bibnamefont{{Abbott}}},
  \bibinfo{author}{\bibfnamefont{F.}~\bibnamefont{{Acernese}}},
  \bibinfo{author}{\bibfnamefont{K.}~\bibnamefont{{Ackley}}},
  \bibinfo{author}{\bibfnamefont{C.}~\bibnamefont{{Adams}}},
  \bibinfo{author}{\bibfnamefont{T.}~\bibnamefont{{Adams}}},
  \bibinfo{author}{\bibfnamefont{P.}~\bibnamefont{{Addesso}}},
  \bibinfo{author}{\bibfnamefont{R.~X.} \bibnamefont{{Adhikari}}},
  \bibinfo{author}{\bibfnamefont{V.~B.} \bibnamefont{{Adya}}},
  \bibnamefont{et~al.}, \bibinfo{journal}{Physical Review Letters}
  \textbf{\bibinfo{volume}{119}}, \bibinfo{eid}{141101}
  (\bibinfo{year}{2017}{\natexlab{a}}), \eprint{1709.09660}.

\bibitem[{\citenamefont{{Abbott}
  et~al.}(2017{\natexlab{b}})\citenamefont{{Abbott}, {Abbott}, {Abbott},
  {Acernese}, {Ackley}, {Adams}, {Adams}, {Addesso}, {Adhikari}, {Adya}
  et~al.}}]{LIGO:2017b}
\bibinfo{author}{\bibfnamefont{B.~P.} \bibnamefont{{Abbott}}},
  \bibinfo{author}{\bibfnamefont{R.}~\bibnamefont{{Abbott}}},
  \bibinfo{author}{\bibfnamefont{T.~D.} \bibnamefont{{Abbott}}},
  \bibinfo{author}{\bibfnamefont{F.}~\bibnamefont{{Acernese}}},
  \bibinfo{author}{\bibfnamefont{K.}~\bibnamefont{{Ackley}}},
  \bibinfo{author}{\bibfnamefont{C.}~\bibnamefont{{Adams}}},
  \bibinfo{author}{\bibfnamefont{T.}~\bibnamefont{{Adams}}},
  \bibinfo{author}{\bibfnamefont{P.}~\bibnamefont{{Addesso}}},
  \bibinfo{author}{\bibfnamefont{R.~X.} \bibnamefont{{Adhikari}}},
  \bibinfo{author}{\bibfnamefont{V.~B.} \bibnamefont{{Adya}}},
  \bibnamefont{et~al.}, \bibinfo{journal}{The Astrophysical Journal Letters}
  \textbf{\bibinfo{volume}{851}}, \bibinfo{eid}{L35}
  (\bibinfo{year}{2017}{\natexlab{b}}).

\bibitem[{\citenamefont{{Abbott}
  et~al.}(2017{\natexlab{c}})\citenamefont{{Abbott}, {Abbott}, {Abbott},
  {Acernese}, {Ackley}, {Adams}, {Adams}, {Addesso}, {Adhikari}, {Adya}
  et~al.}}]{LIGO:2017c}
\bibinfo{author}{\bibfnamefont{B.~P.} \bibnamefont{{Abbott}}},
  \bibinfo{author}{\bibfnamefont{R.}~\bibnamefont{{Abbott}}},
  \bibinfo{author}{\bibfnamefont{T.~D.} \bibnamefont{{Abbott}}},
  \bibinfo{author}{\bibfnamefont{F.}~\bibnamefont{{Acernese}}},
  \bibinfo{author}{\bibfnamefont{K.}~\bibnamefont{{Ackley}}},
  \bibinfo{author}{\bibfnamefont{C.}~\bibnamefont{{Adams}}},
  \bibinfo{author}{\bibfnamefont{T.}~\bibnamefont{{Adams}}},
  \bibinfo{author}{\bibfnamefont{P.}~\bibnamefont{{Addesso}}},
  \bibinfo{author}{\bibfnamefont{R.~X.} \bibnamefont{{Adhikari}}},
  \bibinfo{author}{\bibfnamefont{V.~B.} \bibnamefont{{Adya}}},
  \bibnamefont{et~al.}, \bibinfo{journal}{Physical Review Letters}
  \textbf{\bibinfo{volume}{119}}, \bibinfo{eid}{161101}
  (\bibinfo{year}{2017}{\natexlab{c}}), \eprint{1710.05832}.

\bibitem[{\citenamefont{{Abbott}
  et~al.}(2017{\natexlab{d}})\citenamefont{{Abbott}, {Abbott}, {Abbott},
  {Acernese}, {Ackley}, {Adams}, {Adams}, {Addesso}, {Adhikari}, {Adya}
  et~al.}}]{LIGO:2017e}
\bibinfo{author}{\bibfnamefont{B.~P.} \bibnamefont{{Abbott}}},
  \bibinfo{author}{\bibfnamefont{R.}~\bibnamefont{{Abbott}}},
  \bibinfo{author}{\bibfnamefont{T.~D.} \bibnamefont{{Abbott}}},
  \bibinfo{author}{\bibfnamefont{F.}~\bibnamefont{{Acernese}}},
  \bibinfo{author}{\bibfnamefont{K.}~\bibnamefont{{Ackley}}},
  \bibinfo{author}{\bibfnamefont{C.}~\bibnamefont{{Adams}}},
  \bibinfo{author}{\bibfnamefont{T.}~\bibnamefont{{Adams}}},
  \bibinfo{author}{\bibfnamefont{P.}~\bibnamefont{{Addesso}}},
  \bibinfo{author}{\bibfnamefont{R.~X.} \bibnamefont{{Adhikari}}},
  \bibinfo{author}{\bibfnamefont{V.~B.} \bibnamefont{{Adya}}},
  \bibnamefont{et~al.}, \bibinfo{journal}{The Astrophysical Journal Letters}
  \textbf{\bibinfo{volume}{848}}, \bibinfo{pages}{L13}
  (\bibinfo{year}{2017}{\natexlab{d}}), \eprint{1710.05834}.

\bibitem[{\citenamefont{{Abbott}
  et~al.}(2017{\natexlab{e}})\citenamefont{{Abbott}, {Abbott}, {Abbott},
  {Acernese}, {Ackley}, {Adams}, {Adams}, {Addesso}, {Adhikari}, {Adya}
  et~al.}}]{LIGO:2017d}
\bibinfo{author}{\bibfnamefont{B.~P.} \bibnamefont{{Abbott}}},
  \bibinfo{author}{\bibfnamefont{R.}~\bibnamefont{{Abbott}}},
  \bibinfo{author}{\bibfnamefont{T.~D.} \bibnamefont{{Abbott}}},
  \bibinfo{author}{\bibfnamefont{F.}~\bibnamefont{{Acernese}}},
  \bibinfo{author}{\bibfnamefont{K.}~\bibnamefont{{Ackley}}},
  \bibinfo{author}{\bibfnamefont{C.}~\bibnamefont{{Adams}}},
  \bibinfo{author}{\bibfnamefont{T.}~\bibnamefont{{Adams}}},
  \bibinfo{author}{\bibfnamefont{P.}~\bibnamefont{{Addesso}}},
  \bibinfo{author}{\bibfnamefont{R.~X.} \bibnamefont{{Adhikari}}},
  \bibinfo{author}{\bibfnamefont{V.~B.} \bibnamefont{{Adya}}},
  \bibnamefont{et~al.}, \bibinfo{journal}{The Astrophysical Journal Letters}
  \textbf{\bibinfo{volume}{848}}, \bibinfo{eid}{L12}
  (\bibinfo{year}{2017}{\natexlab{e}}), \eprint{1710.05833}.

\bibitem[{sim(2018)}]{simflowny_webpage}
\emph{\bibinfo{title}{Simflowny project website.}} (\bibinfo{year}{2018}),
  \urlprefix\url{https://bitbucket.org/iac3/simflowny/overview}.

\bibitem[{\citenamefont{Arbona et~al.}(2013)\citenamefont{Arbona, Artigues,
  Bona-Casas, Mass\'o, Mi{\~n}ano, Rigo, Trias, and Bona}}]{Arbona:2013}
\bibinfo{author}{\bibfnamefont{A.}~\bibnamefont{Arbona}},
  \bibinfo{author}{\bibfnamefont{A.}~\bibnamefont{Artigues}},
  \bibinfo{author}{\bibfnamefont{C.}~\bibnamefont{Bona-Casas}},
  \bibinfo{author}{\bibfnamefont{J.}~\bibnamefont{Mass\'o}},
  \bibinfo{author}{\bibfnamefont{B.}~\bibnamefont{Mi{\~n}ano}},
  \bibinfo{author}{\bibfnamefont{A.}~\bibnamefont{Rigo}},
  \bibinfo{author}{\bibfnamefont{M.}~\bibnamefont{Trias}}, \bibnamefont{and}
  \bibinfo{author}{\bibfnamefont{C.}~\bibnamefont{Bona}},
  \bibinfo{journal}{Computer Physics Communications}
  \textbf{\bibinfo{volume}{184}}, \bibinfo{pages}{2321 }
  (\bibinfo{year}{2013}), ISSN \bibinfo{issn}{0010-4655},
  \urlprefix\url{http://www.sciencedirect.com/science/article/pii/S0010465513001471}.

\bibitem[{\citenamefont{Arbona et~al.}(2018)\citenamefont{Arbona, Miñano,
  Rigo, Bona, Palenzuela, Artigues, Bona-Casas, and Massó}}]{Arbona:2017}
\bibinfo{author}{\bibfnamefont{A.}~\bibnamefont{Arbona}},
  \bibinfo{author}{\bibfnamefont{B.}~\bibnamefont{Miñano}},
  \bibinfo{author}{\bibfnamefont{A.}~\bibnamefont{Rigo}},
  \bibinfo{author}{\bibfnamefont{C.}~\bibnamefont{Bona}},
  \bibinfo{author}{\bibfnamefont{C.}~\bibnamefont{Palenzuela}},
  \bibinfo{author}{\bibfnamefont{A.}~\bibnamefont{Artigues}},
  \bibinfo{author}{\bibfnamefont{C.}~\bibnamefont{Bona-Casas}},
  \bibnamefont{and} \bibinfo{author}{\bibfnamefont{J.}~\bibnamefont{Massó}},
  \bibinfo{journal}{Computer Physics Communications}
  \textbf{\bibinfo{volume}{229}}, \bibinfo{pages}{170 } (\bibinfo{year}{2018}),
  ISSN \bibinfo{issn}{0010-4655},
  \urlprefix\url{http://www.sciencedirect.com/science/article/pii/S0010465518300870}.

\bibitem[{\citenamefont{Fritzson et~al.}(2006)\citenamefont{Fritzson, Aronsson,
  Pop, Lundvall, Nystrom, Saldamli, Broman, and Sandholm}}]{openmodelica}
\bibinfo{author}{\bibfnamefont{P.}~\bibnamefont{Fritzson}},
  \bibinfo{author}{\bibfnamefont{P.}~\bibnamefont{Aronsson}},
  \bibinfo{author}{\bibfnamefont{A.}~\bibnamefont{Pop}},
  \bibinfo{author}{\bibfnamefont{H.}~\bibnamefont{Lundvall}},
  \bibinfo{author}{\bibfnamefont{K.}~\bibnamefont{Nystrom}},
  \bibinfo{author}{\bibfnamefont{L.}~\bibnamefont{Saldamli}},
  \bibinfo{author}{\bibfnamefont{D.}~\bibnamefont{Broman}}, \bibnamefont{and}
  \bibinfo{author}{\bibfnamefont{A.}~\bibnamefont{Sandholm}}, in
  \emph{\bibinfo{booktitle}{Computer Aided Control System Design, 2006 IEEE
  International Conference on Control Applications, 2006 IEEE International
  Symposium on Intelligent Control, 2006 IEEE}} (\bibinfo{organization}{IEEE},
  \bibinfo{year}{2006}), pp. \bibinfo{pages}{1588--1595}.

\bibitem[{\citenamefont{Jasak et~al.}(2007)\citenamefont{Jasak, Jemcov, and
  Tokovi\'{c}}}]{openfoam}
\bibinfo{author}{\bibfnamefont{H.}~\bibnamefont{Jasak}},
  \bibinfo{author}{\bibfnamefont{A.}~\bibnamefont{Jemcov}}, \bibnamefont{and}
  \bibinfo{author}{\bibfnamefont{v.}~\bibnamefont{Tokovi\'{c}}}, in
  \emph{\bibinfo{booktitle}{International Workshop on Coupled Methods in
  Numerical Dynamics}} (\bibinfo{year}{2007}), IUC,
  \urlprefix\url{http://cmnd2007.fsb.hr/proc/jasak.pdf}.

\bibitem[{\citenamefont{Davies et~al.}(2011)\citenamefont{Davies, Wilson, and
  Kramer}}]{fluidity}
\bibinfo{author}{\bibfnamefont{D.~R.} \bibnamefont{Davies}},
  \bibinfo{author}{\bibfnamefont{C.~R.} \bibnamefont{Wilson}},
  \bibnamefont{and} \bibinfo{author}{\bibfnamefont{S.~C.}
  \bibnamefont{Kramer}}, \bibinfo{journal}{Geochemistry, Geophysics,
  Geosystems} \textbf{\bibinfo{volume}{12}} (\bibinfo{year}{2011}).

\bibitem[{\citenamefont{Logg et~al.}(2012)\citenamefont{Logg, Mardal, and
  Wells}}]{fenics}
\bibinfo{author}{\bibfnamefont{A.}~\bibnamefont{Logg}},
  \bibinfo{author}{\bibfnamefont{K.-A.} \bibnamefont{Mardal}},
  \bibnamefont{and} \bibinfo{author}{\bibfnamefont{G.}~\bibnamefont{Wells}},
  \emph{\bibinfo{title}{Automated solution of differential equations by the
  finite element method: The FEniCS book}}, vol.~\bibinfo{volume}{84}
  (\bibinfo{publisher}{Springer Science \& Business Media},
  \bibinfo{year}{2012}).

\bibitem[{\citenamefont{Rathgeber et~al.}(2016)\citenamefont{Rathgeber, Ham,
  Mitchell, Lange, Luporini, McRae, Bercea, Markall, and Kelly}}]{firedrake}
\bibinfo{author}{\bibfnamefont{F.}~\bibnamefont{Rathgeber}},
  \bibinfo{author}{\bibfnamefont{D.~A.} \bibnamefont{Ham}},
  \bibinfo{author}{\bibfnamefont{L.}~\bibnamefont{Mitchell}},
  \bibinfo{author}{\bibfnamefont{M.}~\bibnamefont{Lange}},
  \bibinfo{author}{\bibfnamefont{F.}~\bibnamefont{Luporini}},
  \bibinfo{author}{\bibfnamefont{A.~T.} \bibnamefont{McRae}},
  \bibinfo{author}{\bibfnamefont{G.-T.} \bibnamefont{Bercea}},
  \bibinfo{author}{\bibfnamefont{G.~R.} \bibnamefont{Markall}},
  \bibnamefont{and} \bibinfo{author}{\bibfnamefont{P.~H.} \bibnamefont{Kelly}},
  \bibinfo{journal}{ACM Transactions on Mathematical Software (TOMS)}
  \textbf{\bibinfo{volume}{43}}, \bibinfo{pages}{24} (\bibinfo{year}{2016}).

\bibitem[{\citenamefont{Schnetter et~al.}(2015)\citenamefont{Schnetter,
  Blazewicz, Brandt, Koppelman, and L{\"o}ffler}}]{chemora}
\bibinfo{author}{\bibfnamefont{E.}~\bibnamefont{Schnetter}},
  \bibinfo{author}{\bibfnamefont{M.}~\bibnamefont{Blazewicz}},
  \bibinfo{author}{\bibfnamefont{S.~R.} \bibnamefont{Brandt}},
  \bibinfo{author}{\bibfnamefont{D.~M.} \bibnamefont{Koppelman}},
  \bibnamefont{and}
  \bibinfo{author}{\bibfnamefont{F.}~\bibnamefont{L{\"o}ffler}},
  \bibinfo{journal}{Computing in Science Engineering}
  \textbf{\bibinfo{volume}{17}}, \bibinfo{pages}{53} (\bibinfo{year}{2015}).

\bibitem[{\citenamefont{Goodale et~al.}(2003)\citenamefont{Goodale, Allen,
  Lanfermann, Mass\'{o}, Radke, Seidel, and Shalf}}]{cactus}
\bibinfo{author}{\bibfnamefont{T.}~\bibnamefont{Goodale}},
  \bibinfo{author}{\bibfnamefont{G.}~\bibnamefont{Allen}},
  \bibinfo{author}{\bibfnamefont{G.}~\bibnamefont{Lanfermann}},
  \bibinfo{author}{\bibfnamefont{J.}~\bibnamefont{Mass\'{o}}},
  \bibinfo{author}{\bibfnamefont{T.}~\bibnamefont{Radke}},
  \bibinfo{author}{\bibfnamefont{E.}~\bibnamefont{Seidel}}, \bibnamefont{and}
  \bibinfo{author}{\bibfnamefont{J.}~\bibnamefont{Shalf}}, in
  \emph{\bibinfo{booktitle}{Proceedings of the 5th international conference on
  High performance computing for computational science}}
  (\bibinfo{publisher}{Springer-Verlag}, \bibinfo{address}{Berlin, Heidelberg},
  \bibinfo{year}{2003}), VECPAR'02, pp. \bibinfo{pages}{197--227}, ISBN
  \bibinfo{isbn}{3-540-00852-7},
  \urlprefix\url{http://dl.acm.org/citation.cfm?id=1766851.1766868}.

\bibitem[{\citenamefont{Hornung and Kohn}(2002)}]{Hornung:2002}
\bibinfo{author}{\bibfnamefont{R.~D.} \bibnamefont{Hornung}} \bibnamefont{and}
  \bibinfo{author}{\bibfnamefont{S.~R.} \bibnamefont{Kohn}},
  \bibinfo{journal}{Concurrency and Computation: Practice and Experience}
  \textbf{\bibinfo{volume}{14}}, \bibinfo{pages}{347} (\bibinfo{year}{2002}),
  ISSN \bibinfo{issn}{1532-0634},
  \urlprefix\url{http://dx.doi.org/10.1002/cpe.652}.

\bibitem[{\citenamefont{Gunney and Anderson}(2016)}]{Gunney:2016}
\bibinfo{author}{\bibfnamefont{B.~T.} \bibnamefont{Gunney}} \bibnamefont{and}
  \bibinfo{author}{\bibfnamefont{R.~W.} \bibnamefont{Anderson}},
  \bibinfo{journal}{Journal of Parallel and Distributed Computing}
  \textbf{\bibinfo{volume}{89}}, \bibinfo{pages}{65 } (\bibinfo{year}{2016}),
  ISSN \bibinfo{issn}{0743-7315},
  \urlprefix\url{http://www.sciencedirect.com/science/article/pii/S0743731515002129}.

\bibitem[{sam(2015)}]{samrai_webpage}
\emph{\bibinfo{title}{Samrai project website.}} (\bibinfo{year}{2015}),
  \urlprefix\url{https://computation.llnl.gov/project/SAMRAI/}.

\bibitem[{\citenamefont{Gustafsson et~al.}(2013)\citenamefont{Gustafsson,
  Kreiss, and Oliger}}]{Gustafsson:2013}
\bibinfo{author}{\bibfnamefont{B.}~\bibnamefont{Gustafsson}},
  \bibinfo{author}{\bibfnamefont{H.-O.} \bibnamefont{Kreiss}},
  \bibnamefont{and} \bibinfo{author}{\bibfnamefont{J.}~\bibnamefont{Oliger}},
  \emph{\bibinfo{title}{Time-Dependent Problems and Difference Methods}}
  (\bibinfo{publisher}{John Wiley and Sons, Inc.}, \bibinfo{year}{2013}).

\bibitem[{\citenamefont{Toro}(1997)}]{Toro:1997}
\bibinfo{author}{\bibfnamefont{E.}~\bibnamefont{Toro}},
  \emph{\bibinfo{title}{Riemann Solvers and Numerical Methods for Fluid
  Dynamics: A Practical Introduction}} (\bibinfo{publisher}{Springer},
  \bibinfo{year}{1997}), ISBN \bibinfo{isbn}{9783540616764},
  \urlprefix\url{https://books.google.es/books?id=6QFAAQAAIAAJ}.

\bibitem[{\citenamefont{{Calabrese} et~al.}(2004)\citenamefont{{Calabrese},
  {Lehner}, {Reula}, {Sarbach}, and {Tiglio}}}]{Calabrese:2004}
\bibinfo{author}{\bibfnamefont{G.}~\bibnamefont{{Calabrese}}},
  \bibinfo{author}{\bibfnamefont{L.}~\bibnamefont{{Lehner}}},
  \bibinfo{author}{\bibfnamefont{O.}~\bibnamefont{{Reula}}},
  \bibinfo{author}{\bibfnamefont{O.}~\bibnamefont{{Sarbach}}},
  \bibnamefont{and} \bibinfo{author}{\bibfnamefont{M.}~\bibnamefont{{Tiglio}}},
  \bibinfo{journal}{Classical and Quantum Gravity}
  \textbf{\bibinfo{volume}{21}}, \bibinfo{pages}{5735} (\bibinfo{year}{2004}),
  \eprint{gr-qc/0308007}.

\bibitem[{\citenamefont{Harten et~al.}(1983)\citenamefont{Harten, Lax, and van
  Leer}}]{Harten:1983}
\bibinfo{author}{\bibfnamefont{A.}~\bibnamefont{Harten}},
  \bibinfo{author}{\bibfnamefont{P.~D.} \bibnamefont{Lax}}, \bibnamefont{and}
  \bibinfo{author}{\bibfnamefont{B.}~\bibnamefont{van Leer}},
  \bibinfo{journal}{SIAM Review} \textbf{\bibinfo{volume}{25}},
  \bibinfo{pages}{35} (\bibinfo{year}{1983}),
  \eprint{https://doi.org/10.1137/1025002},
  \urlprefix\url{https://doi.org/10.1137/1025002}.

\bibitem[{\citenamefont{{Colella} and {Woodward}}(1984)}]{Colella:1984}
\bibinfo{author}{\bibfnamefont{P.}~\bibnamefont{{Colella}}} \bibnamefont{and}
  \bibinfo{author}{\bibfnamefont{P.~R.} \bibnamefont{{Woodward}}},
  \bibinfo{journal}{Journal of Computational Physics}
  \textbf{\bibinfo{volume}{54}}, \bibinfo{pages}{174} (\bibinfo{year}{1984}).

\bibitem[{\citenamefont{Suresh and Huynh}(1997)}]{Suresh:1997}
\bibinfo{author}{\bibfnamefont{A.}~\bibnamefont{Suresh}} \bibnamefont{and}
  \bibinfo{author}{\bibfnamefont{H.}~\bibnamefont{Huynh}},
  \bibinfo{journal}{Journal of Computational Physics}
  \textbf{\bibinfo{volume}{136}}, \bibinfo{pages}{83 } (\bibinfo{year}{1997}),
  ISSN \bibinfo{issn}{0021-9991},
  \urlprefix\url{http://www.sciencedirect.com/science/article/pii/S0021999197957454}.

\bibitem[{\citenamefont{{Bona} et~al.}(2009)\citenamefont{{Bona}, {Bona-Casas},
  and {Terradas}}}]{Bona:2009}
\bibinfo{author}{\bibfnamefont{C.}~\bibnamefont{{Bona}}},
  \bibinfo{author}{\bibfnamefont{C.}~\bibnamefont{{Bona-Casas}}},
  \bibnamefont{and}
  \bibinfo{author}{\bibfnamefont{J.}~\bibnamefont{{Terradas}}},
  \bibinfo{journal}{Journal of Computational Physics}
  \textbf{\bibinfo{volume}{228}}, \bibinfo{pages}{2266} (\bibinfo{year}{2009}),
  \eprint{0810.2185}.

\bibitem[{\citenamefont{Jiang and Shu}(1996)}]{Jiang:1996}
\bibinfo{author}{\bibfnamefont{G.-S.} \bibnamefont{Jiang}} \bibnamefont{and}
  \bibinfo{author}{\bibfnamefont{C.-W.} \bibnamefont{Shu}},
  \bibinfo{journal}{Journal of Computational Physics}
  \textbf{\bibinfo{volume}{126}}, \bibinfo{pages}{202 } (\bibinfo{year}{1996}),
  ISSN \bibinfo{issn}{0021-9991},
  \urlprefix\url{http://www.sciencedirect.com/science/article/pii/S0021999196901308}.

\bibitem[{\citenamefont{Shu}(1998)}]{Shu:1998}
\bibinfo{author}{\bibfnamefont{C.-W.} \bibnamefont{Shu}},
  \emph{\bibinfo{title}{Essentially non-oscillatory and weighted essentially
  non-oscillatory schemes for hyperbolic conservation laws}}
  (\bibinfo{publisher}{Springer Berlin Heidelberg}, \bibinfo{address}{Berlin,
  Heidelberg}, \bibinfo{year}{1998}), pp. \bibinfo{pages}{325--432}, ISBN
  \bibinfo{isbn}{978-3-540-49804-9},
  \urlprefix\url{https://doi.org/10.1007/BFb0096355}.

\bibitem[{\citenamefont{{Balsara}}(2017)}]{Balsara:2017}
\bibinfo{author}{\bibfnamefont{D.~S.} \bibnamefont{{Balsara}}},
  \bibinfo{journal}{Living Reviews in Computational Astrophysics}
  \textbf{\bibinfo{volume}{3}}, \bibinfo{eid}{2} (\bibinfo{year}{2017}),
  \eprint{1703.01241}.

\bibitem[{\citenamefont{Butcher}(2008)}]{Butcher:2008}
\bibinfo{author}{\bibfnamefont{J.~C.} \bibnamefont{Butcher}},
  \emph{\bibinfo{title}{Numerical Methods for Ordinary Differential Equations}}
  (\bibinfo{publisher}{John Wiley and Sons, Ltd}, \bibinfo{year}{2008}), ISBN
  \bibinfo{isbn}{9780470753767},
  \urlprefix\url{http://dx.doi.org/10.1002/9780470753767.fmatter}.

\bibitem[{\citenamefont{Sebastian and Shu}(2003)}]{Sebastian:2003}
\bibinfo{author}{\bibfnamefont{K.}~\bibnamefont{Sebastian}} \bibnamefont{and}
  \bibinfo{author}{\bibfnamefont{C.-W.} \bibnamefont{Shu}},
  \bibinfo{journal}{Journal of Scientific Computing}
  \textbf{\bibinfo{volume}{19}}, \bibinfo{pages}{405} (\bibinfo{year}{2003}),
  ISSN \bibinfo{issn}{1573-7691},
  \urlprefix\url{https://doi.org/10.1023/A:1025372429380}.

\bibitem[{\citenamefont{{Lehner} et~al.}(2006)\citenamefont{{Lehner},
  {Liebling}, and {Reula}}}]{Lehner:2006}
\bibinfo{author}{\bibfnamefont{L.}~\bibnamefont{{Lehner}}},
  \bibinfo{author}{\bibfnamefont{S.~L.} \bibnamefont{{Liebling}}},
  \bibnamefont{and} \bibinfo{author}{\bibfnamefont{O.}~\bibnamefont{{Reula}}},
  \bibinfo{journal}{Classical and Quantum Gravity}
  \textbf{\bibinfo{volume}{23}}, \bibinfo{pages}{S421} (\bibinfo{year}{2006}),
  \eprint{gr-qc/0510111}.

\bibitem[{\citenamefont{McCorquodale and Colella}(2011)}]{McCorquodale:2011}
\bibinfo{author}{\bibfnamefont{P.}~\bibnamefont{McCorquodale}}
  \bibnamefont{and} \bibinfo{author}{\bibfnamefont{P.}~\bibnamefont{Colella}},
  \bibinfo{journal}{Commun. Appl. Math. Comput. Sci.}
  \textbf{\bibinfo{volume}{6}}, \bibinfo{pages}{1} (\bibinfo{year}{2011}),
  \urlprefix\url{https://doi.org/10.2140/camcos.2011.6.1}.

\bibitem[{\citenamefont{{Mongwane}}(2015)}]{Mongwane:2015}
\bibinfo{author}{\bibfnamefont{B.}~\bibnamefont{{Mongwane}}},
  \bibinfo{journal}{General Relativity and Gravitation}
  \textbf{\bibinfo{volume}{47}}, \bibinfo{eid}{60} (\bibinfo{year}{2015}),
  \eprint{1504.07609}.

\bibitem[{\citenamefont{{Dedner} et~al.}(2002)\citenamefont{{Dedner}, {Kemm},
  {Kr{\"o}ner}, {Munz}, {Schnitzer}, and {Wesenberg}}}]{2002JCoPh.175..645D}
\bibinfo{author}{\bibfnamefont{A.}~\bibnamefont{{Dedner}}},
  \bibinfo{author}{\bibfnamefont{F.}~\bibnamefont{{Kemm}}},
  \bibinfo{author}{\bibfnamefont{D.}~\bibnamefont{{Kr{\"o}ner}}},
  \bibinfo{author}{\bibfnamefont{C.-D.} \bibnamefont{{Munz}}},
  \bibinfo{author}{\bibfnamefont{T.}~\bibnamefont{{Schnitzer}}},
  \bibnamefont{and}
  \bibinfo{author}{\bibfnamefont{M.}~\bibnamefont{{Wesenberg}}},
  \bibinfo{journal}{Journal of Computational Physics}
  \textbf{\bibinfo{volume}{175}}, \bibinfo{pages}{645} (\bibinfo{year}{2002}).

\bibitem[{\citenamefont{{T{\'o}th}}(2000)}]{Toth:2000}
\bibinfo{author}{\bibfnamefont{G.}~\bibnamefont{{T{\'o}th}}},
  \bibinfo{journal}{Journal of Computational Physics}
  \textbf{\bibinfo{volume}{161}}, \bibinfo{pages}{605} (\bibinfo{year}{2000}).

\bibitem[{\citenamefont{{Brio} and {Wu}}(1988)}]{Brio:1988}
\bibinfo{author}{\bibfnamefont{M.}~\bibnamefont{{Brio}}} \bibnamefont{and}
  \bibinfo{author}{\bibfnamefont{C.~C.} \bibnamefont{{Wu}}},
  \bibinfo{journal}{Journal of Computational Physics}
  \textbf{\bibinfo{volume}{75}}, \bibinfo{pages}{400} (\bibinfo{year}{1988}).

\bibitem[{\citenamefont{{Del Zanna} et~al.}(2007)\citenamefont{{Del Zanna},
  {Zanotti}, {Bucciantini}, and {Londrillo}}}]{DelZanna:2007}
\bibinfo{author}{\bibfnamefont{L.}~\bibnamefont{{Del Zanna}}},
  \bibinfo{author}{\bibfnamefont{O.}~\bibnamefont{{Zanotti}}},
  \bibinfo{author}{\bibfnamefont{N.}~\bibnamefont{{Bucciantini}}},
  \bibnamefont{and}
  \bibinfo{author}{\bibfnamefont{P.}~\bibnamefont{{Londrillo}}},
  \bibinfo{journal}{Astronomy and Astrophysics} \textbf{\bibinfo{volume}{473}},
  \bibinfo{pages}{11} (\bibinfo{year}{2007}), \eprint{0704.3206}.

\bibitem[{\citenamefont{{Stone} et~al.}(2008)\citenamefont{{Stone}, {Gardiner},
  {Teuben}, {Hawley}, and {Simon}}}]{Stone:2008}
\bibinfo{author}{\bibfnamefont{J.~M.} \bibnamefont{{Stone}}},
  \bibinfo{author}{\bibfnamefont{T.~A.} \bibnamefont{{Gardiner}}},
  \bibinfo{author}{\bibfnamefont{P.}~\bibnamefont{{Teuben}}},
  \bibinfo{author}{\bibfnamefont{J.~F.} \bibnamefont{{Hawley}}},
  \bibnamefont{and} \bibinfo{author}{\bibfnamefont{J.~B.}
  \bibnamefont{{Simon}}}, \bibinfo{journal}{The Astrophysical Journal
  Supplement Series} \textbf{\bibinfo{volume}{178}}, \bibinfo{eid}{137-177}
  (\bibinfo{year}{2008}), \eprint{0804.0402}.

\bibitem[{\citenamefont{{Bona} et~al.}(2003)\citenamefont{{Bona}, {Ledvinka},
  {Palenzuela}, and {{\v Z}{\'a}{\v c}ek}}}]{Bona:2003}
\bibinfo{author}{\bibfnamefont{C.}~\bibnamefont{{Bona}}},
  \bibinfo{author}{\bibfnamefont{T.}~\bibnamefont{{Ledvinka}}},
  \bibinfo{author}{\bibfnamefont{C.}~\bibnamefont{{Palenzuela}}},
  \bibnamefont{and} \bibinfo{author}{\bibfnamefont{M.}~\bibnamefont{{{\v
  Z}{\'a}{\v c}ek}}}, \bibinfo{journal}{\prd} \textbf{\bibinfo{volume}{67}},
  \bibinfo{eid}{104005} (\bibinfo{year}{2003}), \eprint{gr-qc/0302083}.

\bibitem[{\citenamefont{{Alic} et~al.}(2012)\citenamefont{{Alic}, {Bona-Casas},
  {Bona}, {Rezzolla}, and {Palenzuela}}}]{Alic:2012}
\bibinfo{author}{\bibfnamefont{D.}~\bibnamefont{{Alic}}},
  \bibinfo{author}{\bibfnamefont{C.}~\bibnamefont{{Bona-Casas}}},
  \bibinfo{author}{\bibfnamefont{C.}~\bibnamefont{{Bona}}},
  \bibinfo{author}{\bibfnamefont{L.}~\bibnamefont{{Rezzolla}}},
  \bibnamefont{and}
  \bibinfo{author}{\bibfnamefont{C.}~\bibnamefont{{Palenzuela}}},
  \bibinfo{journal}{\prd} \textbf{\bibinfo{volume}{85}}, \bibinfo{eid}{064040}
  (\bibinfo{year}{2012}), \eprint{1106.2254}.

\bibitem[{\citenamefont{{Bezares} et~al.}(2017)\citenamefont{{Bezares},
  {Palenzuela}, and {Bona}}}]{Bezares:2017}
\bibinfo{author}{\bibfnamefont{M.}~\bibnamefont{{Bezares}}},
  \bibinfo{author}{\bibfnamefont{C.}~\bibnamefont{{Palenzuela}}},
  \bibnamefont{and} \bibinfo{author}{\bibfnamefont{C.}~\bibnamefont{{Bona}}},
  \bibinfo{journal}{\prd} \textbf{\bibinfo{volume}{95}}, \bibinfo{eid}{124005}
  (\bibinfo{year}{2017}), \eprint{1705.01071}.

\bibitem[{\citenamefont{{Palenzuela} et~al.}(2017)\citenamefont{{Palenzuela},
  {Pani}, {Bezares}, {Cardoso}, {Lehner}, and {Liebling}}}]{Palenzuela:2017}
\bibinfo{author}{\bibfnamefont{C.}~\bibnamefont{{Palenzuela}}},
  \bibinfo{author}{\bibfnamefont{P.}~\bibnamefont{{Pani}}},
  \bibinfo{author}{\bibfnamefont{M.}~\bibnamefont{{Bezares}}},
  \bibinfo{author}{\bibfnamefont{V.}~\bibnamefont{{Cardoso}}},
  \bibinfo{author}{\bibfnamefont{L.}~\bibnamefont{{Lehner}}}, \bibnamefont{and}
  \bibinfo{author}{\bibfnamefont{S.}~\bibnamefont{{Liebling}}},
  \bibinfo{journal}{\prd} \textbf{\bibinfo{volume}{96}}, \bibinfo{eid}{104058}
  (\bibinfo{year}{2017}), \eprint{1710.09432}.

\bibitem[{\citenamefont{{Bona} et~al.}(1995)\citenamefont{{Bona}, {Mass{\'o}},
  {Seidel}, and {Stela}}}]{Bona:1995}
\bibinfo{author}{\bibfnamefont{C.}~\bibnamefont{{Bona}}},
  \bibinfo{author}{\bibfnamefont{J.}~\bibnamefont{{Mass{\'o}}}},
  \bibinfo{author}{\bibfnamefont{E.}~\bibnamefont{{Seidel}}}, \bibnamefont{and}
  \bibinfo{author}{\bibfnamefont{J.}~\bibnamefont{{Stela}}},
  \bibinfo{journal}{Physical Review Letters} \textbf{\bibinfo{volume}{75}},
  \bibinfo{pages}{600} (\bibinfo{year}{1995}), \eprint{gr-qc/9412071}.

\bibitem[{\citenamefont{{Alcubierre} et~al.}(2003)\citenamefont{{Alcubierre},
  {Br{\"u}gmann}, {Diener}, {Koppitz}, {Pollney}, {Seidel}, and
  {Takahashi}}}]{Alcubierre:2003}
\bibinfo{author}{\bibfnamefont{M.}~\bibnamefont{{Alcubierre}}},
  \bibinfo{author}{\bibfnamefont{B.}~\bibnamefont{{Br{\"u}gmann}}},
  \bibinfo{author}{\bibfnamefont{P.}~\bibnamefont{{Diener}}},
  \bibinfo{author}{\bibfnamefont{M.}~\bibnamefont{{Koppitz}}},
  \bibinfo{author}{\bibfnamefont{D.}~\bibnamefont{{Pollney}}},
  \bibinfo{author}{\bibfnamefont{E.}~\bibnamefont{{Seidel}}}, \bibnamefont{and}
  \bibinfo{author}{\bibfnamefont{R.}~\bibnamefont{{Takahashi}}},
  \bibinfo{journal}{\prd} \textbf{\bibinfo{volume}{67}}, \bibinfo{eid}{084023}
  (\bibinfo{year}{2003}), \eprint{gr-qc/0206072}.

\bibitem[{\citenamefont{Cook}(2000)}]{Cook:2000}
\bibinfo{author}{\bibfnamefont{G.~B.} \bibnamefont{Cook}},
  \bibinfo{journal}{Living Reviews in Relativity} \textbf{\bibinfo{volume}{3}},
  \bibinfo{pages}{5} (\bibinfo{year}{2000}), ISSN \bibinfo{issn}{1433-8351},
  \urlprefix\url{https://doi.org/10.12942/lrr-2000-5}.

\bibitem[{\citenamefont{{Br{\"u}gmann}
  et~al.}(2008)\citenamefont{{Br{\"u}gmann}, {Gonz{\'a}lez}, {Hannam}, {Husa},
  {Sperhake}, and {Tichy}}}]{Brugmann:2008}
\bibinfo{author}{\bibfnamefont{B.}~\bibnamefont{{Br{\"u}gmann}}},
  \bibinfo{author}{\bibfnamefont{J.~A.} \bibnamefont{{Gonz{\'a}lez}}},
  \bibinfo{author}{\bibfnamefont{M.}~\bibnamefont{{Hannam}}},
  \bibinfo{author}{\bibfnamefont{S.}~\bibnamefont{{Husa}}},
  \bibinfo{author}{\bibfnamefont{U.}~\bibnamefont{{Sperhake}}},
  \bibnamefont{and} \bibinfo{author}{\bibfnamefont{W.}~\bibnamefont{{Tichy}}},
  \bibinfo{journal}{\prd} \textbf{\bibinfo{volume}{77}}, \bibinfo{eid}{024027}
  (\bibinfo{year}{2008}), \eprint{gr-qc/0610128}.

\bibitem[{\citenamefont{Bishop and Rezzolla}(2016)}]{Bishop:2016}
\bibinfo{author}{\bibfnamefont{N.~T.} \bibnamefont{Bishop}} \bibnamefont{and}
  \bibinfo{author}{\bibfnamefont{L.}~\bibnamefont{Rezzolla}},
  \bibinfo{journal}{Living Reviews in Relativity}
  \textbf{\bibinfo{volume}{19}}, \bibinfo{pages}{2} (\bibinfo{year}{2016}),
  ISSN \bibinfo{issn}{1433-8351},
  \urlprefix\url{https://doi.org/10.1007/s41114-016-0001-9}.

\bibitem[{\citenamefont{Liebling}(2002)}]{Liebling:2002}
\bibinfo{author}{\bibfnamefont{S.~L.} \bibnamefont{Liebling}},
  \bibinfo{journal}{Phys. Rev. D} \textbf{\bibinfo{volume}{66}},
  \bibinfo{pages}{041703} (\bibinfo{year}{2002}).

\bibitem[{\citenamefont{Borges et~al.}(2008)\citenamefont{Borges, Carmona,
  Costa, and Don}}]{Borges:2008}
\bibinfo{author}{\bibfnamefont{R.}~\bibnamefont{Borges}},
  \bibinfo{author}{\bibfnamefont{M.}~\bibnamefont{Carmona}},
  \bibinfo{author}{\bibfnamefont{B.}~\bibnamefont{Costa}}, \bibnamefont{and}
  \bibinfo{author}{\bibfnamefont{W.~S.} \bibnamefont{Don}},
  \bibinfo{journal}{Journal of Computational Physics}
  \textbf{\bibinfo{volume}{227}}, \bibinfo{pages}{3191 }
  (\bibinfo{year}{2008}), ISSN \bibinfo{issn}{0021-9991},
  \urlprefix\url{http://www.sciencedirect.com/science/article/pii/S0021999107005232}.

\bibitem[{\citenamefont{Hairer et~al.}(1987)\citenamefont{Hairer, Norsett, and
  Wanner}}]{Hairer:1987}
\bibinfo{author}{\bibfnamefont{E.}~\bibnamefont{Hairer}},
  \bibinfo{author}{\bibfnamefont{S.~P.} \bibnamefont{Norsett}},
  \bibnamefont{and} \bibinfo{author}{\bibfnamefont{G.}~\bibnamefont{Wanner}},
  \emph{\bibinfo{title}{Solving Ordinary Differential Equations I}}
  (\bibinfo{publisher}{Springer-Verlag Berlin Heidelberg},
  \bibinfo{year}{1987}), ISBN \bibinfo{isbn}{978-3-662-12607-3}.

\end{thebibliography}
\bibliographystyle{apsrev}



\appendix

\section{Third-order WENO}\label{appW3}

Let us write explicitly the procedure for $k=2$, leading
to a $3^{\rm rd}$-order WENO reconstruction:

\begin{itemize}
  \item The reconstructed values ${}^L U^{(r)}_{i+1/2}$
  and ${}^R U^{(r)}_{i-1/2}$ of k$^{\rm th}$-order accuracy are
  \begin{eqnarray}
    ^L U^{(0)}_{i+1/2} &=& \frac{1}{2} U_{i} + \frac{1}{2} U_{i+1} 
     \nonumber \\
    ^L U^{(1)}_{i+1/2} &=& -\frac{1}{2} U_{i-1} + \frac{3}{2} U_{i} 
     \nonumber \\
    ^R U^{(0)}_{i-1/2} &=& - \frac{1}{2} U_{i+1} 
    + \frac{3}{2} U_{i}
    \nonumber \\
    ^R U^{(1)}_{i-1/2} &=&  \frac{1}{2} U_{i} 
    + \frac{1}{2} U_{i-1}             
  \end{eqnarray}
  Notice that ${}^L U^{(r)}_{i-1/2}$ and ${}^R U^{(r)}_{i+1/2}$ can be obtained by substituting $i$ by $i \pm 1$ in the previous expressions.
  
  \item We find the smooth indicators ${}^L \beta^{(r)}_{i+1/2}$ and ${}^R \beta^{(r)}_{i+1/2}$
  \begin{eqnarray}
    ^L \beta^{(0)}_{i+1/2} &=& (U_{i+1} - U_{i})^2 \nonumber \\
    ^L \beta^{(1)}_{i+1/2} &=& (U_{i} - U_{i-1})^2 \nonumber \\
    ^R \beta^{(0)}_{i-1/2} &=& (U_{i} - U_{i+1})^2 \nonumber \\
    ^R \beta^{(1)}_{i-1/2} &=& (U_{i-1} - U_{i})^2 
  \end{eqnarray}
  Again, ${}^L \beta^{(r)}_{i-1/2}$ and ${}^R \beta^{(r)}_{i+1/2}$ can be 
  obtained by substituting $i$ by $i \pm 1$ in the previous expressions.
  
  \item We find the $3^{\rm rd}$-order reconstruction 
  \begin{eqnarray}
      U^{L}_{i+1/2} &=& \omega^{(0)}_{i+1/2} {}^L U^{(0)}_{i+1/2} 
                     + \omega^{(1)}_{i+1/2} {}^L U^{(1)}_{i+1/2} 
      \nonumber \\
      U^{R}_{i-1/2} &=& {\tilde \omega}^{(0)}_{i-1/2} {}^R U^{(0)}_{i-1/2} 
                    + {\tilde \omega}^{(1)}_{i-1/2} {}^R U^{(1)}_{i-1/2} 
  \end{eqnarray}
  with  weights $\omega^{(r)}_{i+1/2}$ and  ${\tilde \omega}^{(r)}_{i+1/2}$ constructed by using the generic formulas
  \begin{eqnarray}
      \omega^{(r)}_{i+1/2} = \frac{\alpha^{(r)}_{i+1/2}}{\sum_{s=0}^{k-1}\alpha^{(s)}_{i+1/2}} 
      ~,~
      {\tilde \omega}^{(r)}_{i-1/2} = \frac{{\tilde \alpha}^{(r)}_{i-1/2}}{\sum_{s=0}^{k-1} {\tilde \alpha}^{(s)}_{i-1/2}} 
  \end{eqnarray}    
  where
  \begin{eqnarray}
      \alpha^{(0)}_{i+1/2} = \frac{2/3}{(\epsilon + {}^L \beta^{(0)}_{i+1/2} )^2}
      ~,~
      \alpha^{(1)}_{i+1/2} = \frac{1/3}{(\epsilon + {}^L \beta^{(1)}_{i+1/2} )^2}           
      \nonumber \\
      {\tilde \alpha}^{(0)}_{i-1/2} = \frac{1/3}{(\epsilon + {}^R \beta^{(0)}_{i-1/2} )^2}
      ~,~      
      {\tilde \alpha}^{(1)}_{i-1/2} = \frac{2/3}{(\epsilon + {}^R \beta^{(1)}_{i-1/2} )^2}            
      \nonumber 
   \end{eqnarray}    
\end{itemize}
and $\epsilon$ is usually set to a very small number. The nominal expected convergence rate is achieved when $\epsilon= \Delta x^{2}$. 

As it was mentioned before, the reconstructed values from the other cells can be found  by substituting $i$ by $i \pm 1$ in the previous expressions, namely 
\begin{eqnarray}
   U^{L}_{i-1/2} &=& \omega^{(0)}_{i-1/2} {}^L U^{(0)}_{i-1/2} 
                  + \omega^{(1)}_{i-1/2} {}^L U^{(1)}_{i-1/2} 
   \nonumber \\
   U^{R}_{i+1/2} &=& {\tilde \omega}^{(0)}_{i+1/2} {}^R U^{(0)}_{i+1/2} 
                 + {\tilde \omega}^{(1)}_{i+1/2} {}^R U^{(1)}_{i+1/2} 
\end{eqnarray}

\section{Fifth-order WENO}\label{appW5}

Let us write explicitly the procedure for the $5^{\rm th}$-order
WENO, obtained with $k=3$.

\begin{itemize}
  \item The k reconstructed values ${}^L U^{(r)}_{i+1/2}$
  and ${}^R U^{(r)}_{i-1/2}$ of k$^{\rm th}$-order accuracy are,
  \begin{eqnarray}
    ^L U^{(0)}_{i+1/2} &=& \frac{2}{6} U_{i} + \frac{5}{6} U_{i+1} - \frac{1}{6} U_{i+2} 
     \nonumber \\
    ^L U^{(1)}_{i+1/2} &=& -\frac{1}{6} U_{i-1} + \frac{5}{6} U_{i} + \frac{2}{6} U_{i+1}
     \nonumber \\
    ^L U^{(2)}_{i+1/2} &=& \frac{2}{6} U_{i-2} - \frac{7}{6} U_{i-1} + \frac{11}{6} U_{i}       
     \nonumber \\
    ^R U^{(0)}_{i-1/2} &=& \frac{2}{6} U_{i+2}  
    - \frac{7}{6} U_{i+1} + \frac{11}{6} U_{i}
     \nonumber \\
    ^R U^{(1)}_{i-1/2} &=& - \frac{1}{6} U_{i+1} + \frac{5}{6} U_{i} + \frac{2}{6} U_{i-1}         
    \nonumber \\
    ^R U^{(2)}_{i-1/2} &=& \frac{2}{6} U_{i} + \frac{5}{6} U_{i-1} -\frac{1}{6} U_{i-2}    
  \end{eqnarray}
  The ${}^L U^{(r)}_{i-1/2}$ and ${}^R U^{(r)}_{i+1/2}$ can be 
  obtained by substituting $i$ by $i \pm 1$ in the previous expressions.
  
  \item We find the smooth indicators ${}^L \beta^{(r)}_{i+1/2}$ and ${}^R \beta^{(r)}_{i+1/2}$
  \begin{eqnarray}
    ^L \beta^{(0)}_{i+1/2} &=& \frac{13}{12} (U_{i} - 2 U_{i+1} + U_{i+2})^2 \nonumber \\
    &+& \frac{1}{4} (3 U_{i} - 4 U_{i+1} + U_{i+2})^2  \nonumber \\
    ^L \beta^{(1)}_{i+1/2} &=& \frac{13}{12} (U_{i-1} - 2 U_{i} + U_{i+1})^2 \nonumber \\
    &+& \frac{1}{4} (U_{i-1} - U_{i+1})^2  \nonumber \\
    ^L \beta^{(2)}_{i+1/2} &=& \frac{13}{12} (U_{i-2} - 2 U_{i-1} + U_{i})^2 \nonumber \\
    &+& \frac{1}{4} (U_{i-2} - 4 U_{i-1} + 3 U_{i})^2 
     \nonumber \\
    ^R \beta^{(0)}_{i-1/2} &=&  \frac{13}{12} (U_{i+2} - 2 U_{i+1} + U_{i})^2 \nonumber \\
        &+& \frac{1}{4} (U_{i+2} - 4 U_{i+1} + 3 U_{i})^2 
       \nonumber \\
    ^R \beta^{(1)}_{i-1/2} &=& \frac{13}{12} (U_{i+1} - 2 U_{i} + U_{i-1})^2 \nonumber \\
    &+& \frac{1}{4} (U_{i+1} - U_{i-1})^2 
     \nonumber \\
    ^R \beta^{(2)}_{i-1/2} &=& \frac{13}{12} (U_{i} - 2 U_{i-1} + U_{i-2})^2 \nonumber \\
        &+& \frac{1}{4} (3 U_{i} - 4 U_{i-1} + U_{i-2})^2 
  \end{eqnarray}
  Again, ${}^L \beta^{(r)}_{i-1/2}$ and ${}^R \beta^{(r)}_{i+1/2}$ can be 
  obtained by substituting $i$ by $i \pm 1$ in the previous expressions.

  \item We find the $5^{\rm th}$-order reconstruction 
  \begin{eqnarray}
      U^{L}_{i+1/2} &=& \omega^{(0)}_{i+1/2} {}^L U^{(0)}_{i+1/2} 
     + \omega^{(1)}_{i+1/2} {}^L U^{(1)}_{i+1/2} 
     + \omega^{(2)}_{i+1/2} {}^L U^{(2)}_{i+1/2}      
      \nonumber \\
      U^{R}_{i-1/2} &=& {\tilde \omega}^{(0)}_{i-1/2} {}^R U^{(0)}_{i-1/2} 
   + {\tilde \omega}^{(1)}_{i-1/2} {}^R U^{(1)}_{i-1/2} 
   + {\tilde \omega}^{(2)}_{i-1/2} {}^R U^{(2)}_{i-1/2}   
   \nonumber
  \end{eqnarray}
  with  weights $\omega^{(r)}_{i+1/2}$ and  ${\tilde \omega}^{(r)}_{i+1/2}$ constructed by using the generic formulas
  \begin{eqnarray}
      \omega^{(r)}_{i+1/2} = \frac{\alpha^{(r)}_{i+1/2}}{\sum_{s=0}^{k-1}\alpha^{(s)}_{i+1/2}} 
      ~,~
      {\tilde \omega}^{(r)}_{i-1/2} = \frac{{\tilde \alpha}^{(r)}_{i-1/2}}{\sum_{s=0}^{k-1} {\tilde \alpha}^{(s)}_{i-1/2}} 
  \end{eqnarray}    
  by using
  \begin{eqnarray}
      \alpha^{(0)}_{i+1/2} &=& \frac{3/10}{(\epsilon + {}^L \beta^{(0)}_{i+1/2} )^2}
      ~,~
      {\tilde \alpha}^{(0)}_{i-1/2} = \frac{1/10}{(\epsilon + {}^R \beta^{(0)}_{i-1/2} )^2}
      \nonumber \\            
      \alpha^{(1)}_{i+1/2} &=& \frac{6/10}{(\epsilon + {}^L \beta^{(1)}_{i+1/2} )^2}           
      ~,~
      {\tilde \alpha}^{(1)}_{i-1/2} = \frac{6/10}{(\epsilon + {}^R \beta^{(1)}_{i-1/2} )^2} 
      \nonumber \\            
      \alpha^{(2)}_{i+1/2} &=& \frac{1/10}{(\epsilon + {}^L \beta^{(2)}_{i+1/2} )^2}                 
      ~,~      
      {\tilde \alpha}^{(2)}_{i-1/2} = \frac{3/10}{(\epsilon + {}^R \beta^{(2)}_{i-1/2} )^2}                  
     \nonumber       
   \end{eqnarray}    
   where $\epsilon$ is usually set to a very small number. 
\end{itemize}

Notice that we need also the reconstructed values from the other cells
\begin{eqnarray}
    U^{L}_{i-1/2} &=& \omega^{(0)}_{i-1/2} {}^L U^{(0)}_{i-1/2} 
    + \omega^{(1)}_{i-1/2} {}^L U^{(1)}_{i-1/2} 
    + \omega^{(2)}_{i-1/2} {}^L U^{(2)}_{i-1/2}       
    \nonumber \\
    U^{R}_{i+1/2} &=& {\tilde \omega}^{(0)}_{i+1/2} {}^R U^{(0)}_{i+1/2} 
  + {\tilde \omega}^{(1)}_{i+1/2} {}^R U^{(1)}_{i+1/2} 
  + {\tilde \omega}^{(2)}_{i+1/2} {}^R U^{(2)}_{i+1/2}     
  \nonumber
\end{eqnarray}
   by substituting $i$ by $i \pm 1$ in the previous expressions.
      
More recently there have been some improvements on the standard (or JS) WENO. One of them is the so-called WENO-Z~\cite{Borges:2008}, where the weights are changed by using 

\begin{eqnarray}
  \alpha^{(r)}_{i+1/2} &=& d_r \biggl( 1 + \biggl[ \frac{{}^L \tau_{i+1/2}}{\epsilon + {}^L \beta^{(r)}_{i+1/2} } \biggr]^q \biggr)
   ~~~~,~~~~ \nonumber \\
  {\tilde \alpha}^{(r)}_{i-1/2} &=& {\tilde d}_r \biggl( 1 + \biggl[ \frac{{}^R \tau_{i-1/2}}{\epsilon + {}^R \beta^{(r)}_{i-1/2} } \biggr]^q \biggr)
\end{eqnarray}    
where $q$ is a coefficient between $[1,2]$ and with ${}^L \tau_{i+1/2}= |{}^L \beta^{(0)}_{i+1/2} - {}^L \beta^{(2)}_{i+1/2}|$ and ${}^R \tau_{i-1/2}= |{}^R \beta^{(0)}_{i-1/2} - {}^R \beta^{(2)}_{i-1/2}|$. The scheme becomes more dissipative when the parameter $q$ is increased.
WENO-Z is 4th-order near simple smooth critical
points (i.e., where $u'_j=0$) for $q=1$ and
attains the designed 5th-order for $q=2$, at
the price of being more dissipative. For all these variants of fifth-order WENOs, the parameter
$\epsilon$ is usually set to a very small number
and the expected convergence rate is achieved if $\epsilon= \Delta x^{4}$.

\section{Dense output interpolator}\label{appDense}

Notice that a dense ouput interpolator can be constructed by using the sub-steps of the RK~\cite{Hairer:1987}. Its generic form is 
\begin{equation}
   U^{n + \theta} = U^{n} + \sum_{j=1}^{s} {b}_{j}(\theta) {k}_j ~~~,~~~
   \theta = \frac{t - t^n}{t^{n+1} - t^{n}}
\end{equation}
where $b_i(\theta)$ are the coefficients to build
the interpolator for a given RK scheme. Notice that
the $m-$derivative can also be computed from this dense
output interpolator as
\begin{equation}\label{denseoutput_derivative}
   \frac{d^{m}}{dt^{m}} U (t^n + \theta \Delta t) = 
   \frac{1}{h^m} \sum_{j=1}^{s} {k}_j 
   \frac{d^{m}}{d\theta^{m}} {b}_{j}(\theta)  + O(h^{4-m})~~~,
\end{equation}

For the standard RK$(4,4)$, it can be shown that there is a unique $3^{\rm rd}$-order interpolator that can be written as
\begin{eqnarray}
   b_{1}(\theta) &=& \theta - \frac{3}{2} \theta^2
                 + \frac{2}{3} \theta^3
                 ~~,~~
   b_{2}(\theta) = b_3(\theta) = \theta^2 -
                   \frac{2}{3} \theta^3
                  ~~,~~ \nonumber \\                
   b_{4}(\theta) &=& \frac{-1}{2}\theta^2 -
                   \frac{2}{3} \theta^3                 
\end{eqnarray}

There is a $2^{\rm nd}$-order interpolator which also satisfies the Strong Stability Preserving (SSP) condition for the SSP-RK$(3,3)$:
\begin{equation}
   b_{1}(\theta) = \theta - \frac{5}{6} \theta^2
                 ~~,~~
   b_{2}(\theta) =  \frac{1}{6} \theta^2
                  ~~,~~
   b_{3}(\theta) = \frac{4}{6}\theta^2                  
\end{equation}

\section{Berger-Oliger without order reduction (BOR)}\label{App_BOR}

Let us explain in detail the different steps of the 
BOR algorithm~\cite{McCorquodale:2011,Mongwane:2015}, which is the most efficient but not common yet in the area. A direct Taylor expansion of the solution at $t=t^n$ leads to
\begin{equation}
   U_{n+1} = U_{n} + \Delta t \, U'_{n}
   + \frac{1}{2} \Delta t^2 \, U''_{n}
   + \frac{1}{6} \Delta t^3 \, U'''_{n}
   + O(\Delta t^4)
\end{equation}
By performing a similar expansion on the $k_i$ of 
the RK we obtain
\begin{eqnarray}
\label{ks}
   k_1 &=& \Delta t \, U'_{n} \\
   k_2 &=& \Delta t \, U'_{n} 
       + c_2 \Delta t^2 \, U''_{n}
   + \frac{1}{2} c_2^2 \Delta t^3 \left[ U'''_{n} - f_U U''_{n} \right]  \nonumber \\
   k_3 &=& \Delta t \, U'_{n} 
       + c_3 \Delta t^2 \, U''_{n}
    \nonumber \\   
   &+& \frac{1}{2} \Delta t^3 \left[ c_3^2 U'''_{n} - 
    \left(c_3^2 - 2 \sum_{j=1}^{3} a_{3j} c_j\right)  f_U U''_{n}  \right]  \nonumber \\
   k_4 &=& \Delta t \, U'_{n} 
         + c_4 \Delta t^2 \, U''_{n}
    \nonumber \\     
       &+& \Delta t^3 \left[ \frac{1}{2} c_4^2 U'''_{n} - 
        \left(\frac{1}{2} c_4^2 - \sum_{j=1}^{4} a_{4j} c_j\right)  f_U U''_{n}  \right] \nonumber 
\end{eqnarray}
where $f_U$ is the Jacobian of $f$. Notice that one could solve
now the derivatives of $U$ in terms of $k_i$. However, the equations
are not linearly independent and it is impossible to solve them.
Instead, we will compute the derivatives here from the
dense output interpolator Eqs.~(\ref{denseoutput_derivative}). Once we have
these derivatives, we can calculate the $k_i$ corresponding to the
RK steps of the fine grid, that is, by doing $\Delta t \rightarrow \Delta t/2$ in Eqs.~(\ref{ks}). From there we can calculate the solution at the different RK sub-steps required for the evolution of the boundary points of the fine grid. Next we will describe in detail the implementation for the two commonly-used RK scheme.

\subsubsection{Standard fourth-order RK}

Let us be more explicit and write down the steps for the standard $4^{\rm th}$-order RK for an arbitrary space resolution ratio $R \equiv \Delta x / \Delta x_F$. First we define the time-step on the fine grid $\Delta t_F = \Delta t/R$. Then we start a loop over the steps on the fine grid, going from $r=0,R-1$ :

\begin{enumerate}

\item define $t^{n+r/R} \equiv t + r \Delta t_F$ and evaluate the solution $U_{n+r/R}(t=t^{n+r/R})$ by using the dense output interpolator.

\item compute $\{U'_{n+r/R} , U''_{n+r/R}, U'''_{n+r/R}, f_U U''_{n+r/R} \}$ from the dense output interpolator as a function of $\{ k_1,k_2, k_3,k_4\}$, that 
is, at $t=t^{n+r/R}$ or $\theta=r/R$. The Jacobian can be obtained directly from the $k_i$ of the coarser grid by computing $f_U U''_{n+r/R} = 4 (k_3 - k_2)/\Delta t^3$.

\item compute $\{ k_1,k_2, k_3,k_4\}$ of the fine grid by using its time-step $\Delta t_F$, namely
\begin{eqnarray}
\label{ks_RK44}
   k_1 &=& \Delta t_F \, U'_{n+r/R} \\
   k_2 &=& \Delta t_F \, U'_{n+r/R} 
       + \frac{1}{2} \Delta t_F^2 \, U''_{n+r/R} 
       \nonumber \\
   &+& \frac{1}{8} \Delta t_F^3 \left[ U'''_{n+r/R} - f_U U''_{n+r/R} \right] \nonumber \\
   k_3 &=& \Delta t_F \, U'_{n+r/R} 
       + \frac{1}{2} \Delta t_F^2 \, U''_{n+r/R} \nonumber \\
   &+& \frac{1}{8} \Delta t_F^3 \left[ U'''_{n+r/R} + f_U U''_{n+r/R}  \right] \nonumber \\
   k_4 &=& \Delta t_F \, U'_{n+r/R} 
         + \Delta t_F^2 \, U''_{n+r/R}
       + \frac{1}{2} \Delta t_F^3 U'''_{n+r/R}  \nonumber
\end{eqnarray}

\item use in each sub-step of the first RK step its intermediate value, that for our RK4 is 

\begin{eqnarray}\label{RK4_implementation3}
   {U}^{(1)} &=& {U}_{n+r/R} \\
   {U}^{(2)} &=& {U}_{n+r/R} + \frac{1}{2} k_{1} \nonumber \\
   {U}^{(3)} &=& {U}_{n+r/R} + \frac{1}{2} k_{2} \nonumber \\
   {U}^{(4)} &=& {U}_{n+r/R} + k_{3} \nonumber \\
   {U}^{n+\frac{r+1}{R}} &=& {U}_{n+r/R} + \frac{1}{6} 
   \left( k_1 + 2 k_2  + 2 k_3 + k_4  \right)  \nonumber
\end{eqnarray} 

\end{enumerate}
The final RK step finalizes at $t^{n+1}$.

\subsubsection{Strong Stability Preserving third-order RK}

Let us write down now the procedure for the SSP $3^{\rm rd}$-order RK for an arbitrary ratio $R$. First we define the time-step on the fine grid $\Delta t_F = \Delta t/R$. Then we start a loop over the steps on the fine grid, going from $r=0,R-1$ :

\begin{enumerate}

\item define $t^{n+r/R} \equiv t + r \Delta t_F$ and evaluate the solution $U_{n+r/R}(t=t^{n+r/R})$ by using the dense output interpolator.

\item compute $\{ U'_{n+r/R} , U''_{n+r/R} \}$ from the dense output interpolator as a function of $\{ k_1,k_2, k_3\}$, that 
is, at $t=t^{n+r/R}$ or $\theta=r/R$. 

\item compute $\{ k_1,k_2, k_3\}$ of the fine grid by using its time-step $\Delta t_F$, namely
\begin{eqnarray}
\label{ks_RK44_2}
   k_1 &=& \Delta t_F \, U'_{n+r/R} \\
   k_2 &=& \Delta t_F \, U'_{n+r/R} 
       + \Delta t_F^2 \, U''_{n+r/R}
 \nonumber \\
   k_3 &=& \Delta t_F \, U'_{n+r/R} 
       + \frac{1}{2} \Delta t_F^2 \, U''_{n+r/R}
       \nonumber
\end{eqnarray}

\item use in each sub-step of the first RK step its intermediate value, that for our RK3 is 
\begin{eqnarray}\label{RK3_implementation}
   {U}^{(1)} &=& {U}_{n+r/R} \\
   {U}^{(2)} &=& {U}_{n+r/R} + k_{1} \nonumber \\
   {U}^{(3)} &=& {U}_{n+r/R} + \frac{1}{4} k_{2}  + \frac{1}{4} k_{3}
   \nonumber \\
   {U}^{n+\frac{r+1}{R}} &=& {U}_{n+r/R} + \frac{1}{6} 
   \left( k_1 + 2 k_2  + 4 k_3  \right)  \nonumber
\end{eqnarray} 

\end{enumerate}
The final RK step finalizes at $t^{n+1}$.

%
%
\end{document}